\documentclass[aps,prd,twocolumn,eqsecnum,showpacs,preprintnumbers,superscriptaddress]{revtex4}
\usepackage{graphicx}

\newcommand{\be}{\begin{equation}}
\newcommand{\ee}{\end{equation}}
\newcommand{\ba}{\begin{eqnarray}}
\newcommand{\ea}{\end{eqnarray}}
\newcommand{\ep}{\varepsilon}
\newcommand{\nn}{\nonumber}

\newcommand{\lra}{\leftrightarrow}

\newcommand{\ita}{\textit}
\newcommand{\shift}{\qquad~~}

\begin{document}


\title{One-loop amplitudes for four-point functions with two external massive
quarks and two external massless partons up to $\mathcal O(\ep^2)$}


\author{J.\ G.\ K\"{o}rner}
\email[Electronic address:]{koerner@thep.physik.uni-mainz.de}
\affiliation{Institut f\"{u}r Physik, Johannes
Gutenberg-Universit\"{a}t, D-55099 Mainz, Germany}

\author{Z.\ Merebashvili}
\email[Electronic address:]{zaza@thep.physik.uni-mainz.de}
\affiliation{Institute of High Energy Physics and Informatization,
Tbilisi State University, 0186 Tbilisi, Georgia}

\author{M.\ Rogal}
\email[Present address: Deutsches Elektronen-Synchrotron DESY, Platanenallee 6, D-15738
       Zeuthen, Germany\\Electronic address:]{Mikhail.Rogal@desy.de}
\affiliation{Institut f\"{u}r Physik, Johannes
Gutenberg-Universit\"{a}t, D-55099 Mainz, Germany}

\date{\today}

\begin{abstract}
We present complete analytical ${\mathcal O}(\ep^2)$ results on the one-loop amplitudes
relevant for the next-to-next-to-leading order (NNLO) quark-parton model description of
the hadroproduction of heavy
quarks as given by the so--called loop-by-loop contributions. All results of the
perturbative calculation are given in the dimensional regularization scheme. These
one-loop amplitudes can also be used  as input in the
determination of the corresponding NNLO cross sections for heavy
flavor photoproduction, and in photon-photon reactions.
\end{abstract}


\maketitle

\section{\label{intro}Introduction}

At the leading order (LO) Born term level, heavy quark
hadroproduction has been studied some time ago \cite{LO:1978}. The
next-to-leading order (NLO) corrections to unpolarized heavy quark hadroproduction
were first presented in
\cite{Nason,Been}, and in \cite{Ellis,Smith} for photoproduction.
Corresponding results with initial particles being longitudinally
polarized were calculated in \cite{Bojak1} and
\cite{Bojaka,Bojakb,MCGa,MCGb}. A calculation of the NLO corrections to
top-quark hadroproduction with spin correlations of the final top quarks was
performed in \cite{Arnd}. Analytical results for the so called ``virtual plus
soft'' terms were presented in \cite{Been,Smith,Bojakb} for the
photoproduction and unpolarized hadroproduction of heavy quarks.
Complete analytic results for the polarized and unpolarized
photoproduction, including real bremsstrahlung, can be found in
\cite{MCGb}.

It is well known that the NLO QCD
predictions for the heavy quark production cross sections suffer from theoretical
errors because of the large uncertainty in
choosing the renormalization and factorization scales. In spite of considerable
progress due to
recent work in bringing closer theory and experiment (see e.g.
\cite{Italians,Hubert}), the need for next-to-next-to-leading order (NNLO)
results for heavy quark production
in QCD is by now clearly understood. The NNLO corrections are expected to
significantly reduce the renormalization and factorization scale dependence inherent
to the NLO parton model predictions.

During the last several years much progress has been achieved in developing and
applying various techniques for an all order resummation of heavy quark
production cross sections in different reactions. This concerns the resummation
of the divergent terms in some specific regions of phase space (so called large
logarithms) to NLO (NLL logs) and NNLO (NNLL logs) leading
logarithmic accuracy.
We may mention the work on the threshold and recoil resummations \cite{Laenen} of NLL
logs in hadronic collisions.
Much activity was also devoted to the resummation of NNLL threshold logs for
heavy quark
production in $e^+e^-$ (see e.g. the informative review \cite{Beneke}
and references therein) and $\gamma \gamma$ \cite{Penin} reactions.
However, this cannot replace the need of having the
exact NNLO results for obvious reasons. In fact, these resummed results could be
better understood when the exact NNLO results are available.

The full calculation of the NNLO corrections to heavy hadron production at hadron
colliders will be a very difficult task to complete. It involves the calculation
of many Feynman diagrams of many different topologies. It is clear that an undertaking
of this dimension will have to involve the efforts of many theorists. As one example,
take the recent two-loop calculation of the heavy quark vertex form
factor \cite{Bernreuther} which can be taken as one of the building blocks of the
NNLO calculation. Another building block are the so--called NNLO loop-by-loop
contributions which we have begun to calculate. The necessary ${\cal O}(\ep^2)$
one-loop
scalar master integrals that enter the calculation have been determined by us in
\cite{KMR}.
The present paper is devoted to the determination of the corresponding ${\cal
O}(\ep^2)$
gluon-- and quark--induced one-loop amplitudes including the full spin and color
content
of the problem. In a sequel to this paper we shall present results
on the square of the one-loop amplitudes thereby completing the calculation of the
loop-by-loop part needed for the description of NNLO heavy hadron production.

\begin{figure*}
\includegraphics[height=4.0cm]{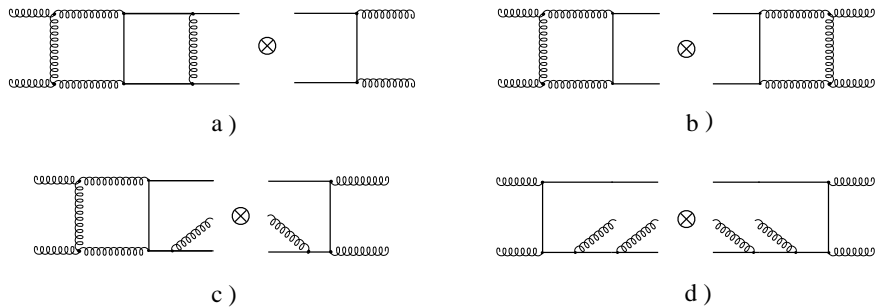}
\caption{\label{fig:exmpl} Exemplary gluon fusion diagrams for the
NNLO calculation of heavy hadron production.}
\end{figure*}

In Fig.~\ref{fig:exmpl} we show one generic diagram each for the
four classes of contributions that need to be calculated for the
NNLO corrections to the gluon--initiated hadroproduction of heavy flavors.
They involve
the two-loop contribution (\ref{fig:exmpl}a), the loop-by-loop
contribution (\ref{fig:exmpl}b), the one-loop gluon emission
contribution (\ref{fig:exmpl}c) and, finally, the two gluon emission
contribution (\ref{fig:exmpl}d). The corresponding graphs for the quark--initiated
processes are not displayed.

In this paper we concentrate on the loop-by-loop contributions exemplified by
Fig.~\ref{fig:exmpl}b.
Specifically, working in the framework of the
dimensional regularization scheme \cite{DREG}, we shall present
${\cal O}(\ep^2)$ results on the one-loop amplitudes.
The expansion of the one--loop amplitudes up to $\ep^2$ is needed
because the one-loop integrals exhibit ultraviolet (UV) and infrared
(IR)/collinear (or mass(M)) singularities up to ${\cal O}(\ep^{-2})$.
When squaring the one-loop amplitudes to obtain the singular and
finite parts of the loop-by-loop
contributions one must thus know the one-loop amplitudes up to $\ep^2$.

In dimensional
regularization there are three different sources that can contribute positive
$\ep$--powers to the Laurent series of the one--loop amplitudes. First, one has the
Laurent series expansion of the scalar one--loop integrals which have been calculated
up to ${\cal O}(\ep^2)$ in \cite{KMR}. Second, the evaluation of the spin algebra
of the
loop amplitudes brings in the $n$--dimensional metric contraction
$g_{\mu \nu} g^{\mu \nu} = n = 4-2\ep$. Third and last, the Passarino-Veltman
decomposition of tensor integrals will again bring in the metric contraction
$g_{\mu \nu} g^{\mu \nu} = n = 4-2\ep$. The latter two points will be treated in
this
paper. It is clear that through the interplay of the three different sources of
positive
$\ep$--powers the Laurent series of the one--loop amplitude itself will, order by
order, contain different orders of the Laurent series coefficient of the scalar
integrals.

We have confirmed the results on the Laurent expansion of the one--loop
amplitude up to ${\cal O}(\ep^0)$ presented in \cite{KM}. These results will not be
listed again in this paper. In this paper we present analytical results
for the coefficients of the $\ep$-- and $\ep^2$--terms of the $\ep$-expansion
including also
their imaginary parts. When presenting our results, we shall
make use of our notation for the coefficient functions of the relevant
scalar integrals calculated up to ${\cal O}(\ep^2)$ in \cite{KMR}.
For the calculation
of the one--loop diagrams with two external massive
quarks and two external massless partons one needs
one scalar one--point function $A$, five scalar two--point functions $B_i$,
six scalar three--point functions $C_i$, and three scalar four-point functions $D_i$.
For example, for the scalar four-point functions $D_i$ we defined successive
coefficient functions $D_i^{(j)}$ according to the expansion
\ba
\nn
\label{Dexp}
&D_i=i C_\ep(m^2)\Big\{\frac{1}{\ep^2}D_i^{(-2)} +\frac{1}{\ep}D_i^{(-1)} + D_i^{(0)}
+ \ep D_i^{(1)} \\
&+ \ep^2 D_i^{(2)} + {\mathcal O}(\ep^3) \Big\}\, ,
\ea
where $C_\ep(m^2)$ is defined by
\be
\label{ceps}
C_{\ep}(m^2)\equiv\frac{\Gamma(1+\ep)}{(4\pi)^2}
\left(\frac{4\pi\mu^2}{m^2}\right)^\ep .
\ee
Similar expansions hold for the scalar one--point function $A$, the scalar two--point
functions $B_i$ and the scalar three--point functions $C_i$.
For the convenience of the reader we have included a table from \cite{KMR} where all
the necessary one-loop master scalar integrals are listed.
\begin{table*}
\caption{List of one-, two-, three- and four-point massive one-loop functions
calculated in our previous paper \cite{KMR} up to ${\cal O}(\ep^2)$.}

\begin{tabular}{lclcccl}    \hline\hline
& &Nomenclature of \cite{Been} & Our nomenclature & Novelty && Comments
\\  \hline
1-point & &       $A(m)$     &  $A$  & -- && Re  \\
\hline
2-point & & $B(p_4-p_2,0,m)$ & $B_1$ & -- && Re  \\
        & & $B(p_3+p_4,m,m)$ & $B_2$ & -- && Re, Im   \\
        & & $B(p_4,0,m)$     & $B_3$ & -- && Re   \\
        & & $B(p_2,m,m)$     & $B_4$ & -- && Re   \\
        & &  $B(p_3+p_4,0,0)$ & $B_5$ & -- && Re, Im   \\
\hline
3-point & & $C(p_4,p_3,0,m,0)$ & $C_1$ & new && Re, Im   \\
        & & $C(p_4,-p_2,0,m,m)$ & $C_2$ & new && Re  \\
        & & $C(-p_2,p_4,0,0,m)$ & $C_3$ & -- && Re  \\
        & & $C(-p_2,-p_1,0,0,0)$ & $C_4$ & -- && Re, Im  \\
        & & $C(-p_2,-p_1,m,m,m)$ & $C_5$ & -- && Re, Im  \\
        & & $C(p_3,p_4,m,0,m)$ & $C_6$ & -- && Re, Im  \\
\hline
4-point & & $D(p_4,-p_2,-p_1,0,m,m,m)$ & $D_1$ & new && Re, Im   \\
        & & $D(-p_2,p_4,p_3,0,0,m,0)$ & $D_2$ & new && Re, Im   \\
        & & $D(-p_2,p_4,-p_1,0,0,m,m)$ & $D_3$ & new && Re  \\ \hline\hline
\end{tabular}
\label{t:tab1}
\end{table*}
We note that for the one-loop scalar integrals the UV and IR/M singularities never overlap,
i.e. do not multiply each other. Singularities of order $\ep^{-2}$ appear only when both
IR and M poles are present simultaneously. This last case is realized when the massless
gluon is attached to either massless fermion or a gluon line in the Feynman diagrams.
Consequently, graphs (\ref{fig:ggnlot}a1), (\ref{fig:ggnlot}c1), (\ref{fig:ggnlos}f1)
and (\ref{fig:ggnlos}f2) shown in the next section have only $\ep^{-1}$ poles, while
graphs (\ref{fig:ggnlot}a2), (\ref{fig:ggnlot}a3), (\ref{fig:ggnlot}c3) and
(\ref{fig:ggnlos}g2) have $\ep^{-2}$ poles.
The details of the pole structure of the various Feynman diagrams can be
found in \cite{KM}.

As remarked on before we have endeavoured to calculate the loop-by-loop contributions
in three steps starting with the scalar one--loop integrals, then calculating the
one--loop amplitudes and finally squaring the one--loop amplitudes. If one's interest
is only in the unpolarized rate one can directly move from step 1 to step 3 without
the interim step of having to evaluate the one--loop amplitudes. However, in the
latter case one loses the information on the spin content of the one-loop
contributions
which cannot be reconstructed from the rate expressions. On the other
hand, having expressions for the one--loop amplitudes allows one to easily
derive the one-loop contributions to partonic cross section
including any polarization of the incoming or outgoing particles.
Our results on the one--loop amplitudes are given separately for every Feynman
diagram in order to facilitate the use of the results for other
relevant processes that differ by color factors.

The hadroproduction of heavy flavors proceeds through the following
two partonic channels:
\begin{equation}
\label{gluglu} g + g \rightarrow Q + \overline Q,
\end{equation}
where $g$ denotes a gluon and $Q (\overline {Q})$ denotes a heavy
quark (antiquark), and
\begin{equation}
\label{qbarq} q + \bar{q} \rightarrow Q + \overline Q,
\end{equation}
where $q (\bar{q})$ is a light massless quark (antiquark).

Note that the Abelian part of the NLO result for (\ref{gluglu})
provides the NLO corrections to heavy flavor production by two
on-shell photons
\begin{equation}
\label{gamgam} \gamma + \gamma \rightarrow Q + \overline Q,
\end{equation}
with the appropriate color factor substitutions. The results for
(\ref{gluglu}) can also be used to determine the corresponding
amplitudes for heavy flavor photoproduction
\begin{equation}
\label{gamglu} \gamma + g \rightarrow Q + \overline Q.
\end{equation}
We mention that the partonic processes (\ref{gluglu}) and
(\ref{qbarq}) are needed for the calculation of the contributions of
single- and double-resolved photons in the photonic processes
(\ref{gamgam}) and (\ref{gamglu}).

NLO cross sections for the process (\ref{gamgam}) have been
determined in \cite{Mirkes,Drees,KMC} for unpolarized and in
\cite{KMC,JT} for polarized initial photons. Note that the authors
of \cite{JT} used a nondimensional regularization scheme to
regularize the poles of divergent integrals. In the papers
\cite{Mirkes,JT} analytic results were presented for ``virtual plus
soft'' contributions alone. We also note that complete analytical
results including hard gluon contributions can be found only in
\cite{KMC}. The two--photon reaction (\ref{gamgam}) will be investigated at
future linear colliders. NLO corrections for the heavy quark
production cross section (\ref{gamgam}) with incident on-shell photons
in definite helicity states are of interest in
themselves as they represent an irreducible background to the
intermediate Higgs boson searches for Higgs masses in the range of
90 to 160 GeV (see e.g. \cite{KMC,JT} and references therein).

The paper is organized as follows. Section~\ref{gluonvertex} contains an
outline of our general approach as well as one--loop amplitudes for the
gluon fusion subprocess for the self-energy and vertex contributions
including their renormalization. In Section~III we discuss the
one-loop contributions to the four box diagrams in the same
gluon-gluon subprocess and give a detailed description of our global
checks on gauge invariance for our results. Section~IV presents
analytic results on the quark-antiquark subprocess (\ref{qbarq}).
Our main results are summarized in Section~V. Finally, in two
appendices we present results for the various coefficient functions that appear
in the main text.

\section{\label{gluonvertex}
CONTRIBUTIONS OF THE TWO- AND THREE-POINT FUNCTIONS TO GLUON FUSION
}

\begin{figure*}
\includegraphics{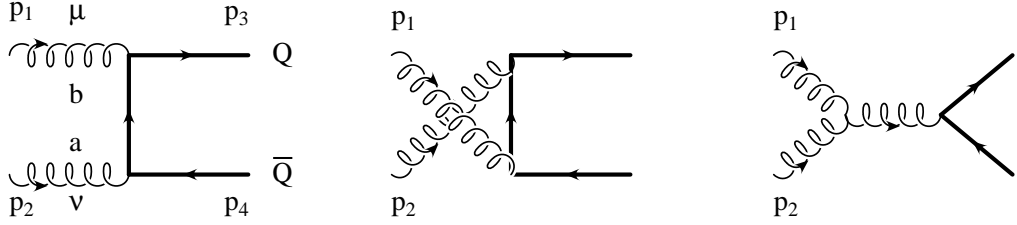}
\caption{\label{fig:gglo}
The $t$-, $u$- and $s$-shannel leading order
(Born) graphs contributing to the gluon (curly lines) fusion
amplitude. The thick solid lines correspond to the heavy quarks.}
\end{figure*}

\begin{figure*}
\includegraphics{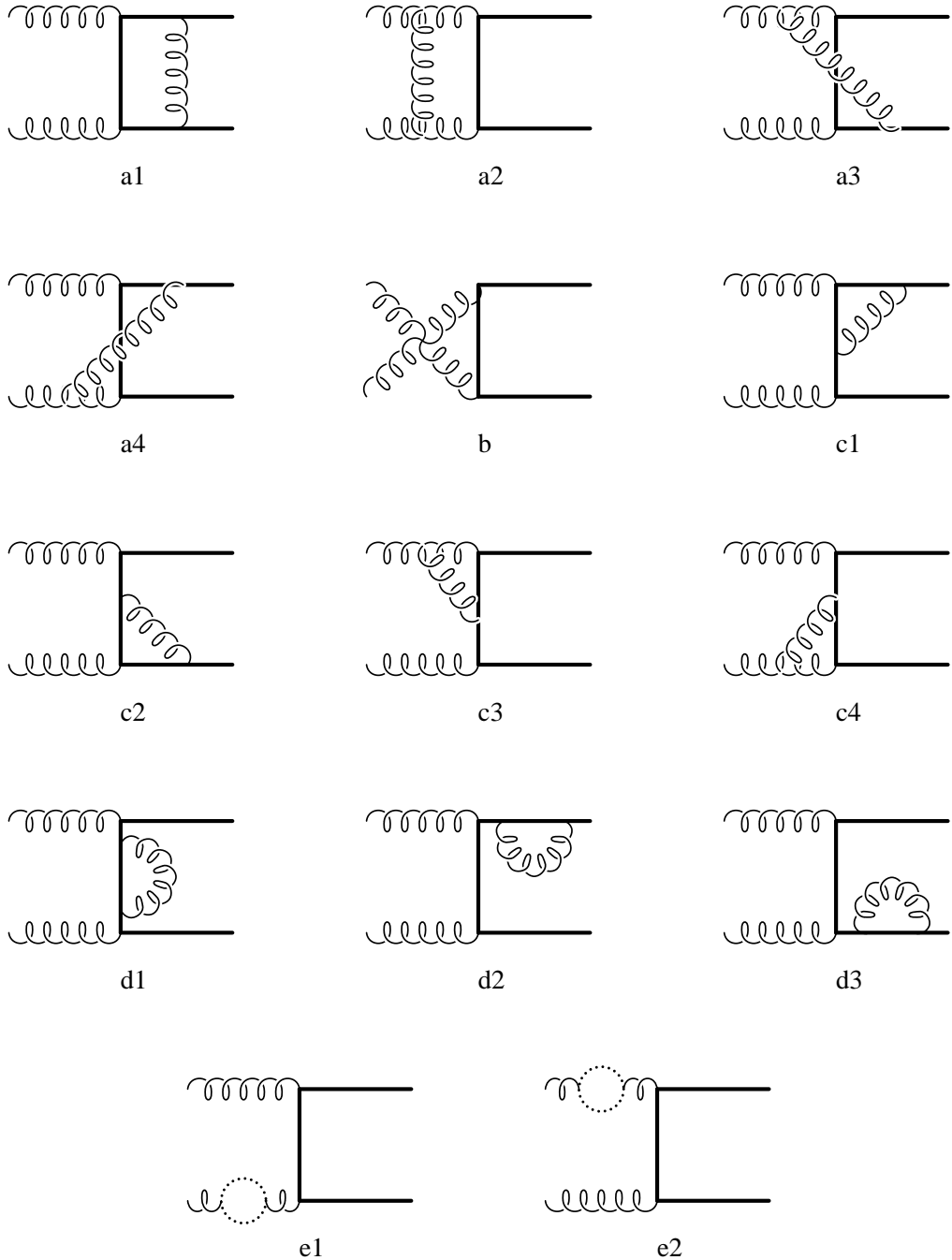}
\caption{\label{fig:ggnlot}
The $t$-channel one-loop graphs contributing to the gluon fusion amplitude.
Loops with dotted lines represent gluon, ghost and light and heavy quarks.}
\end{figure*}

\begin{figure*}
\includegraphics{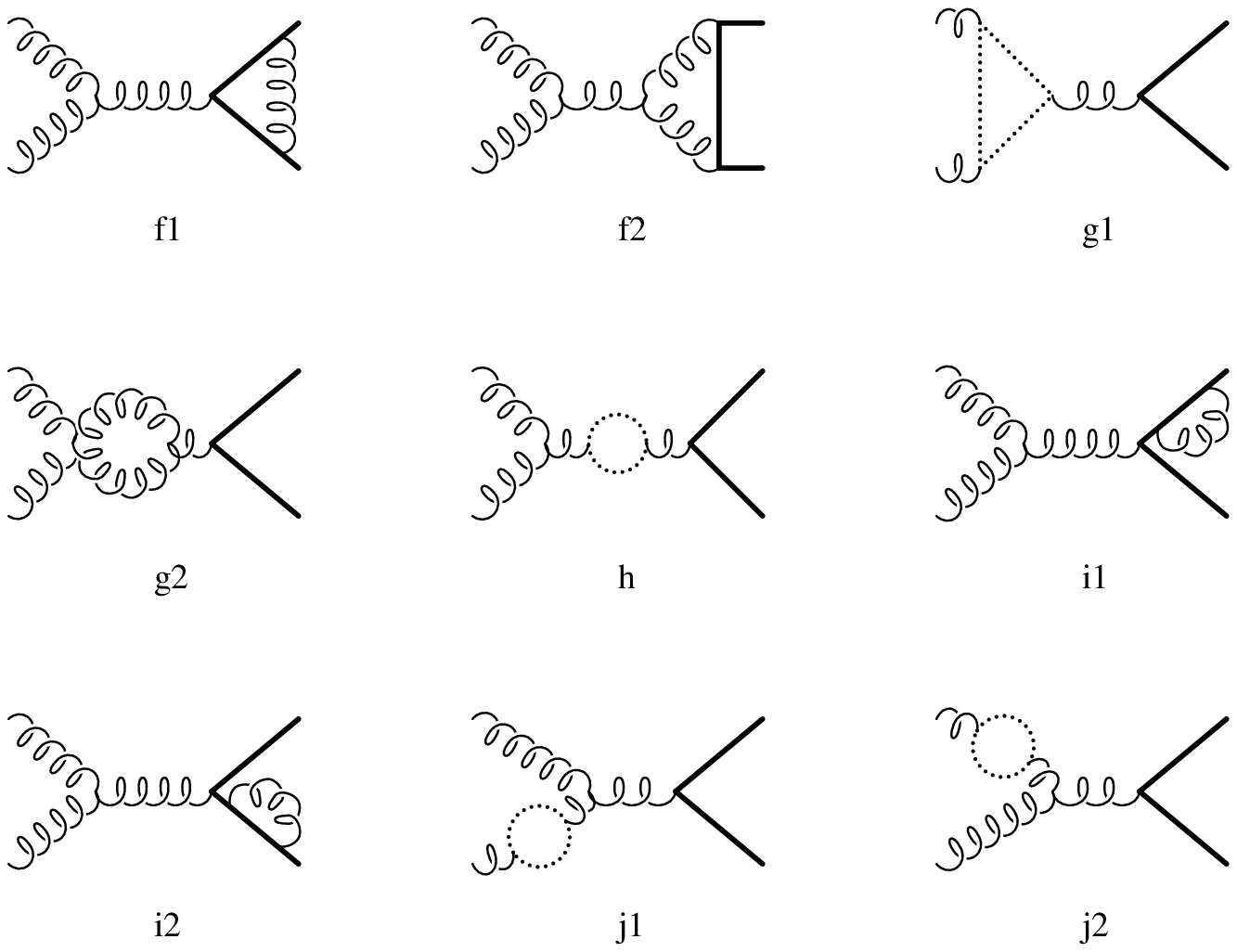}
\caption{\label{fig:ggnlos}
The $s$-channel one-loop graphs contributing to the gluon fusion amplitude.
Loops with dotted lines as in g1,h,j1 and j2 represent gluon, ghost and light and
heavy quarks. The four-gluon coupling contribution appears in g2.}
\end{figure*}

The Born and the one-loop contributions to the partonic gluon fusion
reaction $g(p_1)+g(p_2)\rightarrow Q(p_3) + \overline {Q}(p_4)$ are
shown in Figs.~\ref{fig:gglo}--\ref{fig:ggnlos}. In this section we discuss our
evaluation of the self-energy and vertex  graphs that contribute to
the above subprocess. With the 4-momenta $p_i \, (i=1,...,4)$ as
shown in Fig.~\ref{fig:gglo} and with $m$ the heavy quark mass
we define:
\ba
\nn
&s\equiv (p_1+p_2)^2, \qquad  t\equiv T-m^2 \equiv
(p_1-p_3)^2-m^2,&
\\
&u\equiv U-m^2\equiv (p_2-p_3)^2-m^2.&
\ea

In order to isolate ultraviolet (UV) and infrared/collinear (IR/M)
divergences we have carried out all our calculations in
the dimensional
regularization scheme (DREG) \cite{DREG}
with the dimension of space-time
being formally $n=4-2\ep$.

First of all we note that in general the amplitudes for all the
Feynman diagrams in the gluon fusion subprocess can be written in the
form
\be
M = \epsilon_{\mu}(p_1) \epsilon_{\nu}(p_2) \bar{u}(p_3)
M^{\mu\nu} v(p_4),
\ee
For purposes of brevity, we will present our results in
terms of truncated amplitudes $M^{\mu\nu}$ where the polarization
vectors and Dirac spinors are omitted. For reasons of brevity we shall also refer
to the truncated amplitudes as amplitudes. Of course, the presence of the
polarization vectors and Dirac spinors is implicitly
understood throughout this paper in that the mass shell conditions
$p_1^{\mu}\epsilon_{\mu}(p_1)=0$ and $\not {\rm \hspace{-.03in}} p_3
u(p_3)=m u(p_3)$ etc. are being used to simplify $M^{\mu\nu}$
\footnote{According to the discussion in \cite{Slaven} this implies
that, when further processing our LO and one-loop results in cross
section calculations by folding in the appropriate amplitudes, one
may use the Feynman gauge for the spin sums of polarization vectors.
At the same time, ghost contributions associated with external gluons
have to be omitted.}. Furthermore, $M^{\mu\nu}$ contains the common factor
$C_{\ep}(m^2)$ defined in Eq.~(\ref{ceps}) which arises from
the scalar one-loop integrations described in \cite{KMR}.
Throughout the paper we will omit from all
our one-loop $M^{\mu\nu}$ amplitudes the common factor \be
\label{common} \mathcal C = g^4 C_{\ep}(m^2), \ee where $g$ is the
renormalized coupling constant.

There are three sets of contributing graphs: The $t$--channel, $u$--channel and
the $s$--channel graphs as exemplified in Fig.~\ref{fig:gglo} for the LO Born term
contributions. Since the $u$--channel amplitudes ${\mathcal M}_u$ can be obtained
from the $t$--channel amplitudes ${\mathcal M}_t$ by the relation
\be
\label{tu}
{\mathcal M}_t \leftrightarrow {\mathcal M}_u \equiv
\{     a\leftrightarrow b,  \quad
p_1\leftrightarrow p_2, \quad \mu\leftrightarrow \nu \},
\ee
we shall not list results of the $u$--channel contributions. In (\ref{tu}) $a,b$
are the color indices of the two gluons.
We make it clear from the outset that additional $u$-channel graphs
are obtained from the relevant $t$-channel graphs by the
interchange of the two external bosonic lines (not only momenta).
In exception are the two vertex insertion diagrams (3c3) and (3c4) which will be
discussed later on.
All three interchanges (color, Lorentz indices and bosonic momenta)
have to be done {\it simultaneously}. Note that the second
interchange in (\ref{tu}) implies also the interchange $t\lra u$.
In general, when speaking about the $t$-$u$ symmetry of a given subset of
amplitudes, we
will imply invariance of those amplitudes under the transformations
(\ref{tu}).

We start by writing down amplitudes for the leading order Born
terms. For the $t$-channel gluon fusion subprocess (first graph in
Fig.~\ref{fig:gglo}) we have:
\[
B_t^{\mu\nu} = -i T^b T^a \gamma^{\mu}(  {\rm \hspace{-.08in}} \not
p_3 - {\rm \hspace{-.08in}}  \not p_1 + m) \gamma^{\nu}   /t,
\]
where $T^b$ and $T^a$ are generators ($T^a=\lambda^a/2$, $a=1,...,8$
and the $\lambda^a$ are the usual Gell-Mann matrices) that define the
fundamental representation of the Lie algebra of the color SU(3)
group. Analogously, for the $u$- and $s$-channels depicted in the second and third
graph of Fig.~\ref{fig:gglo} we have, respectively,
\ba
\nn
B_u^{\mu\nu} &=& -i T^a T^b \gamma^{\nu}( {\rm \hspace{-.08in}}
\not p_3
              - {\rm \hspace{-.08in}}  \not p_2 + m)   \gamma^{\mu} /u,   \\
\nn B_s^{\mu\nu} &=& i(T^a T^b - T^b T^a) C_3^{\mu\nu\sigma}
\gamma_{\sigma}/s,
\ea
where the tensor $C_3^{\mu\nu\sigma}$ is obtained from the
Feynman rules for the three-gluon coupling and is given by
\be
C_3^{\mu\nu\sigma} = g_{\mu\nu}(p_1-p_2)_\sigma+g_{\nu\sigma}(p_1+2p_2)_\mu-
                     g_{\mu\sigma}(2p_1+p_2)_\nu.
\ee
We have omitted a common
factor $g^2$ in the Born amplitudes. Acting with Dirac spinors $\bar{u}(p_3)$ and
$v(p_4)$ on the above truncated Born amplitudes from the left and the right,
respectively,
and using the effective relations $p_1^{\mu}=p_2^{\nu}=0$, as remarked on
before, we arrive at the following expressions for the leading order
amplitudes:
\ba
\nn
B_t^{\mu\nu} &=& i T^b T^a ( \gamma^{\mu} {\rm \hspace{-.08in}} \not
p_1
              \gamma^{\nu} - 2 p_3^{\mu} \gamma^{\nu} )/t;          \\
\nn B_u^{\mu\nu}&=& i T^a T^b ( \gamma^{\nu} {\rm \hspace{-.08in}} \not p_2
              \gamma^{\mu} - 2 p_3^{\nu} \gamma^{\mu} )/u;    \\
\nonumber B_s^{\mu\nu} &=& 2i(T^a T^b - T^b T^a)(g^{\mu\nu} {\rm
\hspace{-.1in}}
           \not p_1 + p_2^{\mu} \gamma^{\nu} - p_1^{\nu} \gamma^{\mu})/s.
\ea

Next we proceed with the description of the two-point insertions
to the amplitudes of the subprocess (\ref{gluglu}). But before
we turn to the two-point functions one should mention that our
choice of renormalization scheme will be a \ita{fixed flavor} scheme
throughout this paper. This implies that we have a total number of
flavors $n_f=n_{lf}+1$, where $n_{lf}$ is the number of light (i.e.
massless) flavors and the $``1"$ stands for the produced heavy flavor.
Thus there will only be $n_{lf}$
light flavors involved/active in the $\beta$ function for the
running a QCD coupling $\alpha_s$, and in the splitting functions
that determine the evolution of the structure functions. When having
massless particles in the loops we are using the standard $\overline
{\rm MS}$ scheme, while the contribution of a heavy quark loop in
the gluon self-energy with on-shell external legs is subtracted out
entirely.

Consider first the two $t$-channel self-energy insertion graphs (3d2) and (3d3) in
Fig.~3 with external legs on-shell.
These graphs are very important as they determine the
renormalization parameters in the quark sector. Throughout this
paper we use the so called on-shell prescription for the
renormalization of heavy quarks, the essential ingredients of which
we describe in the following. When dealing with massive quarks one
has to choose a parameter to which one renormalizes the heavy quark
mass. It is natural to choose a quark pole mass for such a parameter
-- the only ``stable'' mass parameter in QCD. The condition on the
renormalized heavy quark self-energy $\Sigma_{r}(\not p)$ is
\be
\label{massren}
\Sigma_{r}(\not p)|_{\not p=m} = 0,
\ee
which removes the singular internal propagator in these self-energy insertion
diagrams. This can be seen from the explicit result for the renormalized heavy
quark external self-energy $\Sigma_{r}(\not p)$
e.g. in dimensional regularization scheme:
\ba
\nn
\Sigma_{r}(\not p) = i g^2 \frac{C_F C_\ep(m^2)}{\ep(1-2\ep)}
\left[ \not \hspace{-.01in}p -  m + \not \hspace{-.01in}p \frac{m^2-p^2}{p^2}
\times  \right.  \\
   \left.
\left(1+\frac{m^2-p^2}{2p^2}(1-\ep)\right) - m\frac{m^2-p^2}{p^2}(2-\ep) \right].
\ea
The above condition (\ref{massren})
determines the mass renormalization
constant $Z_m$.
For the wave function renormalization we have used
the usual condition (see e.g. Ref.~\cite{Ellis})
\be
\label{waveren}
\frac{\partial}{\partial{\rm \hspace{-.07in}}\not p} \Sigma_{r}(\not
p)|_{\not p=m} = 0,
\ee
which fully determines the wave function renormalization constant
$Z_2$.
Since the condition (\ref{waveren}) is not mandatory in
general, there is a freedom in determining the constant $Z_2$.
Note that the condition (\ref{waveren}) sets all external heavy quark
self-energy insertion diagrams to zero, thus making the heavy quark case similar
to the massless one in this regard.
Below we list our expressions for the mass and wave function renormalization
constants.

In the DREG scheme we arrive at the result
\be
Z_m = 1 - g^2 C_F C_{\ep}(m^2) \frac{3-2\ep}{\ep (1-2\ep)}
\ee
which can be expanded in $\ep$ to give
\ba
\nn
Z_m &=& 1 - g^2 C_F C_{\ep}(m^2) \left( \frac{3}{\ep} + 4 + 8\ep + 16\ep^2 +
    {\mathcal O}(\ep^3) \right), \\
Z_2 &=& Z_m,
\ea
%
where $C_F$=4/3 and we do not make a
distinction which poles are of ultraviolet or IR/M origin as we did
in \cite{KM}. \ita{After the mass renormalization
procedure is applied} we obtain the final results for the
two self-energy insertion graphs in the DREG scheme
\ba
\label{grd}
M_{\rm (3d2)}^{\mu\nu} &=& M_{\rm (3d3)}^{\mu\nu} = - C_F
B_t^{\mu\nu}
\frac{3-2\ep}{\ep (1-2\ep)}   \\
\nn
   &=& - C_F B_t^{\mu\nu} \left( \frac{3}{\ep} + 4 + 8\ep + 16\ep^2 +
{\mathcal O}(\ep^3) \right).
\ea
%
%

From here on we will present only results for the $\ep$ and
$\ep^2$ order contributions to the amplitudes.

After addition of the mass renormalization counterterm
the contribution of the
quark self-energy insertion graph (3d1) with external legs off-shell reads:
\ba
\label{grd1}
\nn
M_{\rm (3d1)}^{\mu\nu} &=& C_F B_t^{\mu\nu} \sum_{k=1}^{2} \ep^k
\left( - B_1^{(k)} t/T + 4 B_1^{(k)} m^2/t     \right.  \\
\nn  &&
\left. + B_1^{(k-1)} t/T - k 16 m^2/t \right)
\\
\nn
&-& i C_F T^b T^a m \gamma^{\mu}\gamma^{\nu}  \sum_{k=1}^{2} \ep^k
\left( B_1^{(k)}/T + 2 B_1^{(k)}/t   \right. \\
&& \left.
- B_1^{(k-1)}/T - k 8/t \right).
\ea
The coefficients $B_1^{(k)}$ and $B_1^{(k-1)}$ come from the Laurent series expansion
of the scalar two-point function $B_1$ (see Table 1) quite similar to the corresponding
Laurent series expansion of the four-point functions $D_i$ shown in Eq.~(\ref{Dexp}).

The remaining quark self-energy insertion graphs (4i1) and (4i2) with
external on-shell legs are derived in analogy to the ones considered
above:
\be
\label{gri}
M_{\rm (4i1)}^{\mu\nu} = M_{\rm (4i2)}^{\mu\nu} = - C_F B_s^{\mu\nu}
           \left( 8\ep + 16\ep^2 \right),
\ee

Concerning the gluon self-energy insertion graphs (3e1) and (3e2) with
external legs on-shell, the only nonvanishing contributions are those
from heavy quark loops.
They are given by
\begin{equation}
\label{gre} M_{\rm (3e1)}^{\mu\nu} = M_{\rm (3e2)}^{\mu\nu} = -
B_t^{\mu\nu} \frac{1}{\ep} \,\, \frac{2}{3}.
\end{equation}
However, these contributions are explicitly subtracted (together
with the common
factor $C_{\varepsilon}(m^2)$, see Eqs.~(\ref{ceps}) and
(\ref{common})) in the on-shell renormalization prescription.
Therefore, due to the UV counterterm that subtracts
this loop with heavy quarks, there are no finite
contributions to the amplitudes from these self-energy diagrams. However, at
the same time this counterterm introduces the pole terms from the
light quark loop sector that are needed to cancel soft and collinear
poles from the other parts of the amplitude, e.g. from the real
bremsstrahlung part. This indicates that in practice it is very hard
to completely disentangle UV and IR/M poles in heavy flavor
production and in most cases one obtains a mixture of both instead.

For the reasons specified above we present the gauge
field renormalization constant $Z_3$, used for the gluon self-energy
subtraction:
\ba
\nn
Z_3 = 1 + \frac{g^2}{\ep} \left\{ (\frac{5}{3} N_C - \frac{2}{3}
n_{lf}) C_{\ep}(\mu^2) - \frac{2}{3} C_{\ep}(m^2)\right\}
\\
= 1 + \frac{g^2}{\ep}\left\{ (\beta_0 - 2N_C) C_{\ep}(\mu^2) -
\frac{2}{3} C_{\ep}(m^2)\right\}, \quad
\ea
where the QCD beta--function $\beta_0=(11 N_C - 2 n_{lf})/3$
contains only light quarks. $N_C=3$ is the number of colors.
Accordingly, for the coupling contant renormalization we obtain
\be
Z_g =  1 - \frac{g^2}{\ep}\left\{ \frac{\beta_0}{2} C_{\ep}(\mu^2) -
\frac{1}{3} C_{\ep}(m^2)\right\}.
\ee

As was the case for the diagrams (3e1) and (3e2), diagrams (4j1) and (4j2)
also vanish altogether due to the explicit decoupling of the heavy quarks in
our subtraction prescription.
However, instead of
renormalizing separately each Feynman diagram, one can chose
to employ the renormalization group invariance of the cross section
and do only a mass and coupling constant renormalization. In this
case, knowing the results for the gluon self-energy diagrams turns out to be
useful in checking the complete cancellation of UV poles by just
rescaling the coupling constant in the LO terms $g_{\rm
bare}\rightarrow Z_g g$.
One has
\be
\label{grj}
M_{\rm (4j1)}^{\mu\nu} = M_{\rm (4j2)}^{\mu\nu} = - B_s^{\mu\nu}
\frac{1}{\ep} \,\, \frac{2}{3}.
\ee

Finally we arrive at the gluon self-energy insertion graph (4h), which
contains the off-shell gluon self-energy loop that is used for the
derivation of the renormalization constant $Z_3$. We have evaluated
the internal loop in the Feynman gauge.
In our result we show separately the gauge invariant pieces for
gluon plus ghost, light quarks and one heavy quark flow inside the
loop:
\ba                                                  \label{grhgen}
\nn
& M_{\rm (4h)}^{\mu\nu} =
B_s^{\mu\nu} \left\{ \frac{B_5}{iC_{\ep}(m^2)} \left[ - N_C
\frac{n-14+8\ep}{2(3-2\ep)} - n_{lf} \frac{2(1-\ep)}{3-2\ep} \right]
\right. &  \\
      & \left.
- \frac{1}{\ep} \,\, \frac{2}{3} \,\, I   \right\},
%
\quad  &
\ea
with $n=4-2\ep$ in the DREG scheme.
$B_5$ is the two-point integral whose explicit form is given in \cite{KMR}.
We expand the first line of (\ref{grhgen}) in powers
of $\ep$ and find
\ba \label{grh}                                             \nn
M_{\rm (4h)}^{\mu\nu} &=& B_s^{\mu\nu} \left\{ \left[ N_C \left(
\frac{1}{\ep} \,\, \frac{5}{3} + \frac{31}{9} + \ep\left(
\frac{188}{27} - \frac{5}{3}\, \zeta(2) \right)
                        \right.\right.\right.   \\ \nn &&  \left.
+ \ep^2\left( \frac{1132}{81} - \frac{31}{9}\,
\zeta(2) - \frac{10}{3}\, \zeta(3) \right)\right)            \\
\nn && - n_{lf} \left( \frac{1}{\ep} \,\, \frac{2}{3} + \frac{10}{9}
+ \ep \left( \frac{56}{27} - \frac{2}{3}\, \zeta(2) \right) \right. \\
\nn &&                                                 \left.\left.
+ \ep^2\left( \frac{328}{81} - \frac{10}{9}\, \zeta(2) -
\frac{4}{3}\, \zeta(3) \right)\right) \right]
\left( \frac{-s}{m^2} \right)^{-\ep}          \\
&&    \left.
- \frac{1}{\ep} \,\, \frac{2}{3} \,\, I   \right\}, \ea
with
\ba
\label{integ}
%
I &=& 1 + \ep\left[ -\frac{1}{3} +
B_2^{(0)} \frac{3-\beta^2}{2} \right]                   \\
\nn  &&
+ \ep^2 \left[ -\frac{2}{9} - \frac{1}{3} B_2^{(0)} \beta^2
+ B_2^{(1)} \frac{3-\beta^2}{2} \right]       \\
\nn &&
+ \ep^3 \left[ -\frac{4}{27} - \frac{2}{9} B_2^{(0)} \beta^2
- \frac{1}{3} B_2^{(1)} \beta^2
+ B_2^{(2)} \frac{3-\beta^2}{2} \right].
\ea
In (\ref{integ}) we have made use of the definition
\begin{equation}
\label{varbeta}
\beta \equiv \sqrt{1-4m^2/s}.
\end{equation}


Concluding our discussion on the 2-point insertions we remark that
the amplitudes for the \ita{relevant u-channel 2-point
insertion diagrams} can be obtained from Eqs.~(\ref{grd}), (\ref{grd1}) and
(\ref{gre}) by the transformation (\ref{tu}).

Next we discuss the $t$-- and $u$--channel vertex insertions. In
this paper we write down only the $\ep$-- and $\ep^2$--terms of the
Laurent expansion. The terms proportional to $\ep^{-2}$,
$\ep^{-1}$ and $\ep^0$ can be found in \cite{KM}. We begin with the
purely nonabelian graph (3b) with the four--gluon
vertex. The amplitude takes the following form
\begin{widetext}
\begin{eqnarray}
\label{b}
\nn
M_{\rm (3b)}^{\mu\nu} &=& i N_C (  T^b T^a \sum_{k=1}^2 \ep^k  \{
     ( 2 p_3^{\nu} \gamma^{\mu} + p_4^{\nu} \gamma^{\mu} -
       p_3^{\mu} \gamma^{\nu} - 2 p_4^{\mu} \gamma^{\nu} )
           (B_5^{(k)} + 2 C_1^{(k)} m^2 - 4 k )
- m g^{\mu\nu} (2 B_5^{(k)} + 2 B_5^{(k-1)}      \\
\nn
&&
+ 4 C_1^{(k)} m^2 + C_1^{(k-1)} s - 12 k )
+ 3 m \gamma^{\mu}\gamma^{\nu} (2 B_5^{(k)} + C_1^{(k)} s - 8 k )/2  \}
/(s\beta^2)
+  (a\leftrightarrow b, \mu\leftrightarrow\nu)  )      \\
\nn
&+& i\delta^{ab} \sum_{k=1}^2 \ep^k  \{
         ( p_3^{\nu} \gamma^{\mu} - p_4^{\nu} \gamma^{\mu} +
           p_3^{\mu} \gamma^{\nu} - p_4^{\mu} \gamma^{\nu} )
(B_5^{(k)} + 2 C_1^{(k)} m^2 - 4 k )/2       \\
&&
       + m g^{\mu\nu} ( B_5^{(k)} - 2 B_5^{(k-1)} - 4 C_1^{(k)} m^2 +
       3 C_1^{(k)} s/2 - C_1^{(k-1)} s)
\}  /(s\beta^2).
\end{eqnarray}
\end{widetext}
It is easily seen from Eq.~(\ref{b}) that the amplitude for the graph
(3b) is explicitly $t$-$u$
symmetric, as it follows from the geometric topology of this graph. It is thus
important to state that there is no $u$--channel equivalent of graph (3b).

Next we turn to graphs (3c1) and (3c2). As mentioned before, these
diagrams occur also in
other processes such as photoproduction and $\gamma\gamma$ production of
heavy flavors when one or two of the gluons are replaced by photons. For this reason
we also present the corresponding t-channel
color factors for these graphs.
Then it is
straightforward to separate our Dirac structure from the color coefficients
and one can easily deduce the corresponding results for the other processes
involving photons. In order to facilitate this transscription we list
the color factor for both diagrams
(3c1) and (3c2) which turn out to be the same:
\begin{equation}
T_{\rm col}^{\rm (3c1)} = T_{\rm col}^{\rm (3c2)} =
(C_F - \frac{N_C}{2}) T^b T^a = - \frac{1}{6} T^b T^a.
\end{equation}
The complete amplitudes are:
\begin{widetext}
\begin{eqnarray}
\label{c1}
\nonumber
M_{\rm (3c1)}^{\mu\nu} &=& B_t^{\mu\nu} \sum_{k=1}^2 \ep^k \{
         B_1^{(k)} (6 m^2/t + 1) + 2 B_1^{(k-1)} z_t/t
         + 2 C_2^{(k)} m^2 + 4 C_2^{(k-1)} m^2 - k 8 (4 m^2/t + 1)
\} /6     \\
\nn
&+&   i T^b T^a (
p_3^{\mu} \gamma^{\nu} \sum_{k=1}^2 \ep^k \{
               B_1^{(k)} (z_t/t + t/T)
     + B_1^{(k-1)} (2 z_t/t - t/T) + 2 (C_2^{(k)} + 2 C_2^{(k-1)}) m^2
     - k 8 z_t/t \}                   \\
\nonumber
&+& m p_3^{\mu} {\rm \hspace{-.05in}}   \not p_1 \gamma^{\nu}
\sum_{k=1}^2 \ep^k \{ B_1^{(k)} /T - B_1^{(k-1)} (2/t
                   + 1/T) - 2 C_2^{(k-1)} + k 4/t \}       \\
&-& m \gamma^{\mu} \gamma^{\nu} \sum_{k=1}^2 \ep^k \{
B_1^{(k)} + B_1^{(k-1)} + C_2^{(k-1)} t - 6 k \}
  )/(3t),
\end{eqnarray}
\end{widetext}
where we have introduced the abbreviation $z_t \equiv 2 m^2+t$.

For the graph (3c2) we obtain:
\begin{widetext}
\begin{eqnarray}
\label{c2}
\nonumber
M_{\rm (3c2)}^{\mu\nu} &=& B_t^{\mu\nu} \sum_{k=1}^2 \ep^k \{
        B_1^{(k)} (6 m^2/t + 1) + 2 B_1^{(k-1)} z_t/t
        + 2 C_2^{(k)} m^2 + 4 C_2^{(k-1)} m^2 - k 8 (4 m^2/t + 1) \}
                  /6     \\
\nn
&+&   i T^b T^a ( p_4^{\nu} \gamma^{\mu} \sum_{k=1}^2 \ep^k \{
        B_1^{(k)} (-2 m^2/t - 3 + t/T)
        - B_1^{(k-1)} t/T - 2 C_2^{(k)} m^2 + k 8 T/t \}
                             \\
\nonumber
&+& m p_4^{\nu} ( 2 p_3^{\mu} - \gamma^{\mu}
{\rm \hspace{-.1in}}         \not p_1 ) \sum_{k=1}^2 \ep^k \{
      B_1^{(k)} /T - B_1^{(k-1)} (2/t + 1/T) - 2 C_2^{(k-1)} + k 4/t \}  \\
&-& m \gamma^{\mu} \gamma^{\nu} \sum_{k=1}^2 \ep^k \{
      B_1^{(k)} + B_1^{(k-1)} + C_2^{(k-1)} t - 6 k \}
   )/(3 t).
\end{eqnarray}
\end{widetext}

Next we write down the results for graphs (3c3) and (3c4).
The color factors for both diagrams are the same:
\begin{equation}
T_{\rm col}^{\rm (3c3)} = T_{\rm col}^{\rm (3c4)} =
- \frac{N_C}{2} T^b T^a = - \frac{3}{2} T^b T^a.
\end{equation}
We have
\begin{eqnarray}
\label{c3}
\nonumber
M_{\rm (3c3)}^{\mu\nu} &=& 3 B_t^{\mu\nu} \sum_{k=1}^2 \ep^k \{
     - 3 B_1^{(k)} m^2/t - C_3^{(k)} t      \\
\nn
&&            + k 4 (3 m^2/t + 1) \}    \\
\nn
&& + 3 i T^b T^a (
p_3^{\mu} \gamma^{\nu} \sum_{k=1}^2 \ep^k \{
      B_1^{(k)} m^2 (1/T - 2/t)                     \\
\nn
&&
      + B_1^{(k-1)} t/T - C_3^{(k)} t + k 4 z_t/t   \}        \\
\nonumber
&&    + 3 m \gamma^{\mu}\gamma^{\nu} \sum_{k=1}^2 \ep^k \{
      B_1^{(k)}/2 - 2 k  \}
   \\
\nonumber
&& + m p_3^{\mu} {\rm \hspace{-.05in}} \not p_1 \gamma^{\nu}
\sum_{k=1}^2 \ep^k \{
       B_1^{(k)} (2/t - 1/T)    \\
&&     + B_1^{(k-1)}/T - k 8/t  \}  )/t.
\end{eqnarray}
And
\begin{eqnarray}
\label{c4}
\nonumber
M_{\rm (3c4)}^{\mu\nu} &=& 3 B_t^{\mu\nu} \sum_{k=1}^2 \ep^k \{
    - 3 B_1^{(k)} m^2/t - C_3^{(k)} t      \\
\nn
&&                + k 4 (3 m^2/t + 1)   \}   \\
\nn
&& + 3 i T^b T^a (  p_4^{\nu} \gamma^{\mu} \sum_{k=1}^2 \ep^k \{
    B_1^{(k)} m^2 (1/T - 2/t)                        \\
\nn
&&
    + B_1^{(k-1)} (t/T - 2) + C_3^{(k)} t + k 4 (2 m^2/t - 1) \}  \\
\nonumber
&& + 3 m \gamma^{\mu}\gamma^{\nu} \sum_{k=1}^2 \ep^k \{
       B_1^{(k)}/2 - 2 k  \}     \\
\nonumber
&& + m p_4^{\nu} (2 p_3^{\mu} - \gamma^{\mu}    {\rm \hspace{-.1in}}
\not p_1)  \sum_{k=1}^2 \ep^k \{
       B_1^{(k)} (2/t - 1/T)            \\
&&                         + B_1^{(k-1)}/T - k 8/t  \}   )/t.
\end{eqnarray}
The results for the amplitudes of the {\it relevant u-channel vertex insertion
diagrams} are obtained from Eqs.~(\ref{c1}), (\ref{c2}), (\ref{c3}) and
(\ref{c4}) by the transformation (\ref{tu}).
However, there is a subtle
point involved here: we stress that for the graphs (3c3) and (3c4)
the $M_t \leftrightarrow M_u$ transformation (\ref{tu}) transforms the $t$-channel
result of the graph
(3c3) to the $u$-channel result for the graph (3c4), while the $t$-channel
result of (3c4) goes to the $u$-channel result for (3c3). This is important
to keep in mind when dealing with reactions which involve asymmetric set
of graphs as e.g. in the photoproduction of heavy flavors. The reason for this is
that when doing transformation (\ref{tu}) the three-gluon vertex attached to one
of the initial bosonic lines does not stay attached to the same bosonic line.
However, we note that transformation $p_3\lra p_4$ does uniquely relate all the
$t$- and $u$-channel diagrams for the subprocess under consideration.

Next we turn to the remaining $s$-channel graphs shown in Fig.~\ref{fig:ggnlos}.
For all the gluon propagators we work in Feynman gauge.
This set of graphs is purely nonabelian for the QCD type one-loop
corrections. In the case that one wants to replace the gluonic vertex correction
in graph (4f1) by a photonic vertex correction one needs the explicit form of the
color factor for graph (4f1):
\begin{equation}
T_{\rm col}^{\rm (4f1)} = (C_F - \frac{N_C}{2}) ( T^a T^b - T^b T^a )
= - \frac{1}{6} ( T^a T^b - T^b T^a ).
\end{equation}
The amplitude including the color factor is
\begin{eqnarray}
\nonumber
M_{\rm (4f1)}^{\mu\nu} &=& B_s^{\mu\nu} \sum_{k=1}^2 \ep^k \{
3 B_2^{(k)} + 2 B_2^{(k-1)} + C_6^{(k)} s (1 + \beta^2)    \\
\nn
&&    - 16 k \} /6
+ 2 i (T^a T^b - T^b T^a) m
[ - g^{\mu\nu} (s + 2 t)          \\
&&   \nn
- 4 p_3^{\mu} p_4^{\nu}+ 4 p_4^{\mu} p_3^{\nu}]
\sum_{k=1}^2 \ep^k \{
        B_2^{(k)} + 2 B_2^{(k-1)} - 8 k \}          \\
&&
/(6 s^2\beta^2).
\end{eqnarray}

Graph (4f2) contributes as:
\begin{eqnarray}
\nonumber
M_{\rm (4f2)}^{\mu\nu} &=& N_C B_s^{\mu\nu} \sum_{k=1}^2 \ep^k \{
     B_5^{(k)} (8 m^2-s) + 2 C_1^{(k)} m^2 s    \\
\nn
&& - k 16 (5 m^2-s)  \}/ (2 s \beta^2)    \\
\nn
&& + 2 i N_C (T^a T^b - T^b T^a) m
  [ - g^{\mu\nu} (s + 2 t) -      \\
\nn   &&
4 p_3^{\mu} p_4^{\nu} + 4 p_4^{\mu} p_3^{\nu} ] \sum_{k=1}^2 \ep^k \{
B_5^{(k)} (8 m^2 + s)        \\
\nn   &&
- 2 B_5^{(k-1)} s + 6 C_1^{(k)} m^2 s - C_1^{(k-1)} s^2 -   \\
&&
k 4 (12 m^2 - s)  \}          / 2 s^3\beta^4.
\end{eqnarray}

We end our consideration of the vertex insertions for gluonic fusion with the
sum of the two graphs (4g1) and (4g2) which we refer to as the triangle graph
contribution (tri)$\equiv$(4g1)+(4g2).
For the case when one has gluons and ghosts inside the triangle loop we
obtain:
\begin{eqnarray}
\nonumber
M_{\rm (tri)}^{\mu\nu}(g) &=& - 3 N_C ( B_s^{\mu\nu} \sum_{k=1}^2 \ep^k \{
207 B_5^{(k)} + 12 B_5^{(k-1)}                \\
\nn   &&
+ 54 C_4^{(k)} s + 8 k + (k-1) 8 \tilde B_5^{(0)}     \}          \\
\nonumber
&& + 6 i (T^a T^b - T^b T^a) {\rm \hspace{-.06in}}  \not p_1
\sum_{k=1}^2 \ep^k \{ g^{\mu\nu} [
9 B_5^{(k)}            \\
&&    \nn
- 12 B_5^{(k-1)} + 9 C_4^{(k)} s - 8 k
- (k-1) 8 \tilde B_5^{(0)}     ] /s            \\
&&   \nn
  + 8 p_2^{\mu} p_1^{\nu} [3 B_5^{(k-1)} + 2 k
+ (k-1) 2 \tilde B_5^{(0)}]/s^2 \}        \\
&&      ) /324,
\label{glutri}
\end{eqnarray}
where $\tilde B_5^{(0)} = B_5^{(0)} - 4/3$.
When one has light and heavy quarks inside the
loop one has
\begin{eqnarray}
\nonumber
M_{\rm (tri)}^{\mu\nu}(q) &=&  6 n_{lf} ( B_s^{\mu\nu} \sum_{k=1}^2 \ep^k \{
9 B_5^{(k)} - 3 B_5^{(k-1)} - 2 k              \\
\nn   &&
- (k-1) 2 \mathcal B_5^{(0)} \}
- 3 i (T^a T^b - T^b T^a)
                {\rm \hspace{-.06in}}  \not p_1  \times          \\
\nn   &&
          [g^{\mu\nu}/s - 2 p_2^{\mu} p_1^{\nu}/s^2]
\sum_{k=1}^2 \ep^k \{
          3 B_5^{(k-1)} + 5 k        \\
&&
+ (k-1) (5 \tilde B_5^{(0)} + 3) \}  ) /81
\label{qtri}
\end{eqnarray}
where $n_{lf}$ is the number of light flavors in the triangle loop. For the
heavy flavor case one has
\begin{widetext}
\begin{eqnarray}
\label{qmtri}
\nn
M_{\rm (tri)}^{\mu\nu}(Q) &=& 6  ( B_s^{\mu\nu} \sum_{k=1}^2 \ep^k \{
 6 (3 B_2^{(k)} + 2 B_2^{(k-1)} + (k-1) 4 B_2^{(0)}/3) m^2/s +
 9 B_2^{(k)}  - 3 B_2^{(k-1)} - 2 k - 2 (k-1) \tilde B_2^{(0)} \} \\
\nonumber
&& - i (T^a T^b - T^b T^a)    {\rm \hspace{-.06in}}  \not p_1
     [g^{\mu\nu}/s - 2 p_2^{\mu} p_1^{\nu}/s^2]
\sum_{k=1}^2 \ep^k \{
   12 (3 B_2^{(k)} + 2 B_2^{(k-1)} + (k-1) 4 B_2^{(0)}/3) m^2/s        \\
&&   + 18 (C_5^{(k)} + C_5^{(k-1)} + (k-1) C_5^{(0)}) m^2
+ 3 B_2^{(k-1)} + 5 k + (k-1) (5 \tilde B_2^{(0)} + 3)  \} )   /81,
\end{eqnarray}
\end{widetext}
where $\tilde B_2^{(0)} = B_2^{(0)} - 4/3$.
The complete amplitude for the triangle (tri)$\equiv$(4g1)+(4g2) is the sum
of the above three expressions (\ref{glutri}), (\ref{qtri}) and (\ref{qmtri}):
\begin{equation}
\label{tri}
M_{\rm (tri)}^{\mu\nu} = M_{\rm (tri)}^{\mu\nu}(g) +
                  M_{\rm (tri)}^{\mu\nu}(q) + M_{\rm (tri)}^{\mu\nu}(Q).
\end{equation}

In Ref.~\cite{Andrei} one can find general results for the gluon triangle
in any gauge and dimension. We have compared the first two terms in (\ref{tri}) with
the corresponding expressions in Ref.~\cite{Andrei} and found complete agreement.


\section{\label{gluonboxes}
RESULTS FOR THE BOX DIAGRAMS IN GLUON FUSION
}

In this section we describe the technically most involved derivation of
the 4-point massive box diagrams.
The four box graphs (3a1)--(3a4) contributing to the subprocess
$g+g\rightarrow Q+\overline Q$ are depicted
in Fig.~\ref{fig:ggnlot}. We have used
Passarino-Veltman
techniques \cite{passar} to reduce tensor integrals to scalar ones
where the scalar master integrals are taken from our previous publication
\cite{KMR}.

For each of the gluon fusion box diagrams we expand the truncated amplitude
$M^{\mu\nu}$ in terms of a set of 20 Lorentz-Dirac covariants multiplied by the same
number of invariant functions. In the reduction of the Lorentz-Dirac structure to
this basic set of 20 covariants we have been making use of the mass shell conditions
described in Sec.~II. The 20 Lorentz-Dirac covariants are subdivided into
eight subsets according to their Dirac structure. The 20 invariant functions
multiplying the covariants are sorted according to the contributions of a basic set of
functions $f_i^{(k)}$ (called basis functions) related to the scalar master integrals
of \cite{KMR}. The index $i$ runs over the members of the set of basis functions
occuring in a particular graph. The index $k$ denotes the power of $\ep$ which the
basis function multiplies. The basis functions $f_i^{(k)}$ are multiplied by
coefficient functions $b_{in}^{(j)}$ where the index pair $(n,j)$ identifies the
covariant which the coefficient function multiplies. Note that the basis functions
$f_i^{(k)}$ have been defined such that the coefficient functions $b_{in}^{(j)}$
do not depend on the index $k$.
We thus cast the box amplitude into the following universal
form:
\begin{widetext}
\begin{eqnarray}
\label{genbox}
M^{\mu\nu}&=&i T_{\rm col} \,\,  \sum_{k=1}^2 \ep^k
                    \{ M_{\rm Bt}^{\mu\nu} \sum f_i^{(k)}  b_{i1}^{(0)}  \\
\nonumber  &+&
       \not{p}_1 [ g^{\mu\nu} \sum f_i^{(k)}  b_{i1}^{(1)} +
                   p_3^{\mu} p_3^{\nu} \sum f_i^{(k)}  b_{i2}^{(1)} +
                   p_3^{\mu} p_4^{\nu} \sum f_i^{(k)}  b_{i3}^{(1)} +
                   p_4^{\mu} p_3^{\nu} \sum f_i^{(k)}  b_{i4}^{(1)} +
                   p_4^{\mu} p_4^{\nu} \sum f_i^{(k)}  b_{i5}^{(1)} ]   \\
\nonumber  &+&
              \gamma^{\mu} [ p_3^{\nu} \sum f_i^{(k)}  b_{i1}^{(2)} +
                            p_4^{\nu} \sum f_i^{(k)}  b_{i2}^{(2)} ] +
              \gamma^{\nu} [ p_3^{\mu} \sum f_i^{(k)}  b_{i1}^{(3)} +
                            p_4^{\mu} \sum f_i^{(k)}  b_{i2}^{(3)} ] +
              m \gamma^{\mu} \gamma^{\nu} \sum f_i^{(k)}  b_{i1}^{(4)}    \\
\nonumber  &+&
m \gamma^{\mu}{\rm\hspace{-.1in}}\not{p}_{1} [ p_3^{\nu} \sum f_i^{(k)}
               b_{i1}^{(5)} + p_4^{\nu} \sum f_i^{(k)}  b_{i2}^{(5)} ] +
m \gamma^{\nu}{\rm\hspace{-.1in}}\not{p}_{1} [ p_3^{\mu} \sum f_i^{(k)}
               b_{i1}^{(6)} + p_4^{\mu} \sum f_i^{(k)}  b_{i2}^{(6)} ]    \\
\nonumber  &+&
m [ g^{\mu\nu} \sum f_i^{(k)}  b_{i1}^{(7)} +
    p_3^{\mu} p_3^{\nu} \sum f_i^{(k)}  b_{i2}^{(7)} +
    p_3^{\mu} p_4^{\nu} \sum f_i^{(k)}  b_{i3}^{(7)} +
    p_4^{\mu} p_3^{\nu} \sum f_i^{(k)}  b_{i4}^{(7)} +
    p_4^{\mu} p_4^{\nu} \sum f_i^{(k)}  b_{i5}^{(7)} ] \}   \\
\nonumber  &+&           \{{\cal M}_t \leftrightarrow {\cal M}_u\}.
\end{eqnarray}
\end{widetext}
The symbol $\{{\cal M}_t \leftrightarrow {\cal M}_u\}$ at the end of
Eq.(\ref{genbox}) needs to be explained. It has the same meaning as the symbol
${\cal M}_t \leftrightarrow {\cal M}_u$ defined in Eq.(\ref{tu}) except that
diagrams (3a3) and (3a4) are exempted from the sum.
The crossed boxes (\ref{fig:ggnlot}a3) and (\ref{fig:ggnlot}a4) go into each
other under the ${\cal M}_t \leftrightarrow{\cal M}_u$ operation.
More exactly, for each of these diagrams, when
one symmetrically interchanges the two bosonic lines (together with the appended
three-gluon vertex) one arrives at the original box graph topology since
these boxes represent diagrams of the so called non-planar topology.
This becomes even more clear when one interchanges $p_3\lra p_4$: In this
case each of the two crossed box graphs is reflected into itself.

Taking parity into account one has altogether $2 \cdot 2 \cdot 2 \cdot 2/2 = 8$
independent amplitudes and thus eight independent covariants for the process
$g+g\rightarrow Q+\overline Q$ in $n=4$--dimensions. We have made no attempt to
reduce the 20 (plus 3 from the $u$--channel contributions)
covariants in (\ref{genbox}) to a basic set of independent gauge invariant
covariants. In fact, gauge invariance will be checked later on in terms of the expansion
(\ref{genbox}). At any rate, the number of independent gauge invariant covariants
will very likely change going from $n=4$ to a general $n \neq 4$.

Depending on the type of the box graph one has a different number of terms
in the $(i)$ summation in (\ref{genbox}). These numbers as well as the
set of basis functions $f_i^{(k)}$ related to the scalar master integrals are specified
below. The coefficient functions
$b_{in}^{(j)}$ are given in Appendix~A of this paper.

In the expansion (\ref{genbox}) it is convenient to choose one covariant as the
t-channel Born term amplitude structure $M_{\rm Bt}^{\mu\nu}$ (and correspondingly a
$u$--channel Born term amplitude structure).
We define it as
\begin{equation}
M_{\rm Bt}^{\mu\nu} \equiv \gamma^{\mu} (\not{p}_3-\not{p}_1+m) \gamma^{\nu},
\end{equation}
which, when taken between the spin wave functions implying the
effective relations $p_1^{\mu}=0,\,\, p_2^{\nu}=0$, can be written as
\begin{equation}
\label{bt}
M_{\rm Bt}^{\mu\nu} = 2 p_3^{\mu} \gamma^{\nu}  -  \gamma^{\mu}
           {\rm \hspace{-.1in}}               \not{p}_1 \gamma^{\nu}.
\end{equation}

For each of the box diagrams (\ref{fig:ggnlot}a1) and
(\ref{fig:ggnlot}a2) we found
the following empirical relations between the $b_{in}^{(5)}$ and $b_{in}^{(6)}$
coefficient functions:
\begin{equation}
\label{rel1}
b_{i1}^{(6)}=b_{i2}^{(5)},  \qquad   b_{i2}^{(6)}=b_{i1}^{(5)}.
\end{equation}
Because of the relations (\ref{rel1}) 
we will not write
down the results for the $b_{in}^{(6)}$ coefficients in the Appendix~A.

Next we present the color factors and basis functions for the abelian type box
diagram (\ref{fig:ggnlot}a1).
For this graph the sums over $i$ in (\ref{genbox}) run from
1 to 17 for each of the 20 terms.
One has:
\begin{equation}
T_{\rm col}=\frac{1}{4} \delta^{ab} + (C_F - \frac{N_C}{2}) T^b T^a.
\end{equation}
\begin{eqnarray}
\label{fa1}
\nn   &
f_1^{(k)}=B_1^{(k-1)},  \qquad   f_2^{(k)}=B_1^{(k)},  \qquad
f_3^{(k)}=B_2^{(k-1)}, & \\
\nn   &
f_4^{(k)}=B_2^{(k)},    \qquad    f_5^{(k)}=C_2^{(k-1)},  \qquad
f_6^{(k)}=C_2^{(k)},     &  \\
\nn   &
f_7^{(k)}=C_5^{(k-1)},   \qquad   f_8^{(k)}=C_5^{(k)},    \qquad
f_9^{(k)}=C_6^{(k-1)},   &   \\
\nn   &
f_{10}^{(k)}=C_6^{(k)},    \qquad   f_{11}^{(k)}=D_1^{(k-1)},   \qquad
f_{12}^{(k)}=D_1^{(k)},  &   \\
&
f_{13}^{(k)}=k,          &  \\
\nn   &
f_{14}^{(k)}=(k-1) C_2^{(k-2)},   \qquad
f_{15}^{(k)}=(k-1) C_5^{(k-2)},   &  \\
\nn   &
f_{16}^{(k)}=(k-1) C_6^{(k-2)},   \qquad   f_{17}^{(k)}=(k-1) D_1^{(k-2)}.
\end{eqnarray}

The corresponding coefficient functions $b_{in}^{(j)}$ are
listed in Appendix A. Many of the coefficient functions are in fact related to
each other. One has
\be
b_{12\,n}^{(j)} = - t b_{10\,n}^{(j)},   \qquad   j \ne 0.
\ee
And for any given values of $n$ and $j$ one has
\ba   &
b_{11\,n}^{(j)}=-t b_{9 n}^{(j)},   \qquad   b_{15\,n}^{(j)}=s/(2 t)
                                                    b_{14\,n}^{(j)}, &  \\
\nn   &
b_{16\,n}^{(j)}=s \beta^2/(2 z_t) b_{14\,n}^{(j)},  \qquad
b_{17\,n}^{(j)}=-s t \beta^2/(2 z_t) b_{14\,n}^{(j)}.   &
\ea

Further relations are valid for particular sets of the parameters $n,j$:
\ba
\nn
b_{7 n}^{(j)}=s/(2 t) b_{5 n}^{(j)},   \qquad   j=0,1,2,4 ;    \\
b_{7 n}^{(7)}=s/(2 t) b_{5 n}^{(7)},   \qquad   n=1,3,5 ;    \\
\nn
b_{8 n}^{(j)}=s/(2 t) b_{6 n}^{(j)},   \qquad   j=4,5,6,7.
\ea

Because of these relations among the coefficient functions we will write down only the
independent
coefficients $b_{in}^{(j)}$ in Appendix~A.

For the nonabelian box diagram (\ref{fig:ggnlot}a2) the sums over $i$ in
(\ref{genbox}) again run from 1 to 17 for each of the 20 terms in (\ref{genbox}) .
For the color factor we obtain:
\begin{equation}
T_{\rm col}=\frac{1}{4} \delta^{ab} + \frac{N_C}{2} T^b T^a.
\end{equation}
The relevant seventeen basis functions that describe the result of
evaluating the box diagram (\ref{fig:ggnlot}a2) are given by
\begin{eqnarray}
\label{fa2}
\nn   &
f_1^{(k)}=B_1^{(k-1)},  \qquad   f_2^{(k)}=B_1^{(k)},  \qquad
f_3^{(k)}=B_5^{(k-1)}, & \\
\nn   &
f_4^{(k)}=B_5^{(k)},    \qquad    f_5^{(k)}=C_1^{(k-1)},  \qquad
f_6^{(k)}=C_1^{(k)},     &  \\
\nn   &
f_7^{(k)}=C_3^{(k-1)},   \qquad   f_8^{(k)}=C_3^{(k)},    \qquad
f_9^{(k)}=C_4^{(k-1)},   &   \\
\nn   &
f_{10}^{(k)}=C_4^{(k)},    \qquad   f_{11}^{(k)}=D_2^{(k-1)},   \qquad
f_{12}^{(k)}=D_2^{(k)},  &   \\
&
f_{13}^{(k)}=k,          &  \\
\nn   &
f_{14}^{(k)}=(k-1) C_1^{(k-2)},   \qquad
f_{15}^{(k)}=(k-1) C_3^{(k-2)},   &  \\
\nn   &
f_{16}^{(k)}=(k-1) C_4^{(k-2)},   \qquad   f_{17}^{(k)}=(k-1) D_2^{(k-2)}.
\end{eqnarray}

There are five relations between particular coefficients for the box
diagram (\ref{fig:ggnlot}a2), valid for any values of $n$ and $j$:
\be
\label{Gensa2}
b_{9 n}^{(j)}= s/(2 t) b_{7 n}^{(j)},  \qquad
b_{11 n}^{(j)}= - (s/2)  b_{7 n}^{(j)},
\ee
and
\ba
\label{SetIIa2}
\nn  &
b_{14 n}^{(j)}= s z_t/(2 t^2) b_{15 n}^{(j)},  \qquad
b_{16 n}^{(j)}= s/(2 t) b_{15 n}^{(j)},  &  \\
&
b_{17 n}^{(j)}= - (s/2)  b_{15 n}^{(j)}.   &
\ea
In addition, one has two sets of relations that are valid for
the corresponding parts of the expression (\ref{genbox}) for the box
(\ref{fig:ggnlot}a2).
The first set of relations is
\ba
\label{LessGena21}
&
b_{5 n}^{(j)}= s z_t/(2 t^2) b_{7 n}^{(j)}, & \\
\nn  \\
\label{LessGena22}  &
b_{14 n}^{(j)}= 2 b_{5 n}^{(j)},   \qquad
b_{15 n}^{(j)}= 2 b_{7 n}^{(j)},  &   \\
\nn  &
b_{16 n}^{(j)}= 2 b_{9 n}^{(j)},   \qquad
b_{17 n}^{(j)}= 2 b_{11 n}^{(j)},  &
\ea
The above equalities are valid for $j=0,2,3,4,5,6$ and for $j=1,7
\,\,and\,\, n=1$.
Note that in the presence of the set (\ref{SetIIa2}) not all of the
relations
in (\ref{LessGena21}), (\ref{LessGena22}) are independent. Therefore, we
can choose Eq.~(\ref{LessGena21}) and only one relation (e.g. the
second one) out of the four relations in (\ref{LessGena22}) as a set of
independent relations.

The second set of relations is represented by the two equalities that
are identical to the ones of (\ref{Gensa2}), but are valid only for
$j=4,5,6,7$ or for $j=1 \,\,and\,\,n=2,3,4,5$:
\be
\label{Gensa22}
b_{10 n}^{(j)}= s/(2 t) b_{8 n}^{(j)},  \qquad
b_{12 n}^{(j)}= - (s/2)  b_{8 n}^{(j)},
\ee

In the case of the crossed box (\ref{fig:ggnlot}a4) one has twenty basis functions
for each of the terms in
(\ref{genbox}). The color factor for this graph takes the simple form
\begin{equation}
T_{\rm col}=\frac{1}{4} \delta^{ab}.
\end{equation}
The functions $f_i^k$ are defined as follows:
\begin{eqnarray}
\label{fa4}
\nn   &
f_1^{(k)}=B_1^{(k-1)},  \qquad   f_2^{(k)}=B_1^{(k)},  &  \\
\nn   &
f_3^{(k)}=B_{1u}^{(k-1)},
\qquad   f_4^{(k)}=B_{1u}^{(k)} ,  & \\
\nn   &
f_5^{(k)}=C_2^{(k-1)},  \qquad   f_6^{(k)}=C_2^{(k)},     &  \\
\nn   &
f_7^{(k)}=C_{2u}^{(k-1)},   \qquad   f_8^{(k)}=C_{2u}^{(k)},  & \\
\nn   &
f_9^{(k)}=C_3^{(k-1)},   \qquad    f_{10}^{(k)}=C_3^{(k)},   &  \\
\nn   &
f_{11}^{(k)}=C_{3u}^{(k-1)},   \qquad  f_{12}^{(k)}=C_{3u}^{(k)},  &   \\
\nn   &
f_{13}^{(k)}=D_3^{(k-1)},   \qquad   f_{14}^{(k)}=D_3^{(k)},   &  \\
&
f_{15}^{(k)}=k,          &  \\
\nn   &
f_{16}^{(k)}=(k-1) C_2^{(k-2)},   \qquad   f_{17}^{(k)}=(k-1)
                                             C_{2u}^{(k-2)},   &  \\
\nn   &
f_{18}^{(k)}=(k-1) C_3^{(k-2)},   \qquad   f_{19}^{(k)}=(k-1)
                                             C_{3u}^{(k-2)},   &  \\
\nn   &
f_{20}^{(k)}=(k-1) D_3^{(k-2)},
\end{eqnarray}
where the subscript ``u'' is an operational definition prescribing a
$(t\leftrightarrow u)$ interchange in the argument of that function, i.e.
$B_{1u}^{(k)}=B_{1}^{(k)} (t\leftrightarrow u)$.

There are numerous relations between the $b_{in}^{(j)}$
coefficient functions for this diagram. These relations read:

For any value of $n$ and $j$
\ba   &
b_{11\,n}^{(j)}= b_{9 n}^{(j)} u/t,     \qquad      b_{13\,n}^{(j)}= -
                                                       b_{9 n}^{(j)} u, &
\label{Gens1}
\ea
as well as
\ba
\label{Gens2}
&
b_{7 n}^{(j)}= b_{5 n}^{(j)} u/t,      \qquad   j \ne 5;    &  \\
\nn   &
b_{8 n}^{(j)}= b_{6 n}^{(j)} u/t,  \qquad    j \ne 1,2;    &  \\
\nn   &
b_{12\,n}^{(j)}= b_{10\,n}^{(j)} u/t,   \qquad    j = 4,5,6,7;  & \\
\nn   &
b_{12\,n}^{(1)}= b_{10\,n}^{(1)} u/t,   \qquad    n \ne 1;  & \\
\nn   &
b_{14\,n}^{(1)}= - b_{12\,n}^{(1)} t,   \qquad    n \ne 0,2;  & \\
\nn   &
b_{14,2}^{(2)}= - b_{12,2}^{(2)} t.   \qquad  \qquad \qquad &
\ea
Further one has a less general but still very useful relation for any $n$
\be
b_{9 n}^{(j)}= - b_{5 n}^{(j)} t u/(2D+tu),  \qquad  j = 0,3,6,
\label{LessGens}
\ee
with $D=m^2s-tu$.
Equation (\ref{LessGens}) above is also valid for $j=1,7$ and $n=1$.

For the coefficient functions that effectively only multiply the $\ep^2$--terms we
have two sets of relations.
One set is
\ba
\label{SetI}   &
b_{16\,n}^{(j)} = 2 b_{5 n}^{(j)},  & \\
\nn   &
b_{17\,n}^{(j)} = 2 b_{7 n}^{(j)}, \qquad  b_{18\,n}^{(j)} = 2 b_{9
                                                           n}^{(j)},  & \\
\nn   &
b_{19\,n}^{(j)} = 2 b_{11\,n}^{(j)}, \qquad  b_{20\,n}^{(j)} = 2
                                                       b_{13\,n}^{(j)}\, &
\ea
which are valid for the same values
of $j,n$ as specified in and after (\ref{LessGens}).
The other set reads
\ba
\label{SetII}   &
b_{17\,n}^{(j)} = b_{16\,n}^{(j)} u/t,  & \\
\nn  &
b_{18\,n}^{(j)} = - b_{16\,n}^{(j)} t u/(2D+tu),  &  \\
\nn  &
b_{19\,n}^{(j)} = - b_{16\,n}^{(j)} u^2/(2D+tu),  &  \\
\nn  &
b_{20\,n}^{(j)} = b_{16\,n}^{(j)} t u^2/(2D+tu). &
\ea
The relations (\ref{SetII}) are global
for the crossed box (\ref{fig:ggnlot}a4), i.e. valid for any set of index values.
Because the relations (\ref{SetI}) always occur together
with the relations (\ref{LessGens}), only the first
relation of (\ref{SetI}) is important. The other four relations in (\ref{SetI})
are redundant since they can be derived from
(\ref{Gens1}), the first relation in (\ref{Gens2}), (\ref{LessGens})
and (\ref{SetII}).

In addition to the relations listed above, various coefficient functions of the
crossed box are related by $(t \lra u)$ exchange. For instance, the coefficient
functions
multiplying the Born term structure $M_{\rm Bt}^{\mu\nu}$ (or $j=0$) are related
by
\ba
\label{bsym}
\nn  &
b_{i,1}^{(0)}= b_{i+2,1}^{(0)}(t \lra u), &  \quad
i=1,2,5,6,9,10;   \\
&
b_{i,1}^{(0)}= b_{i+1,1}^{(0)}(t \lra u), & \quad  i=16,18;
\ea
The remaining $(j=0)$ coefficient functions turn into themselves under $(t \lra u)$.

\begin{figure*}
\includegraphics{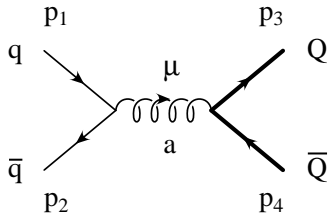}
\caption{\label{fig:qqlo}
The lowest order Feynman diagram contributing to the subprocess
$q \bar{q} \rightarrow Q \overline Q$. The thick lines correspond
to the heavy quarks.}
\end{figure*}

Other coefficient functions are negatively related by $(t\lra u)$--exchange:
\ba
\label{bantisym}
\nn   &
b_{i,4}^{(1)} =- b_{i+2,3}^{(1)}(t\lra u),  &  \quad
                                                i=1,2,5,6,9,10;   \\
\nn   &
b_{i,4}^{(1)}  = - b_{i,3}^{(1)}(t\lra u),  & \quad  i=13,14,15,20;  \\
\nn   &
b_{i,4}^{(1)} = - b_{i+1,3}^{(1)}(t\lra u),  & \quad  i=16,18.   \\
\\
\nn   &
b_{i,5}^{(1)} = - b_{i+2,2}^{(1)}(t\lra u),  &  \quad
                                                    i=1,2,5,6,9,10;   \\
\nn   &
b_{i,5}^{(1)} = - b_{i,2}^{(1)}(t\lra u),  & \quad  i=13,14,15,20;  \\
\nn   &
b_{i,5}^{(1)} = - b_{i+1,2}^{(1)}(t\lra u),  & \quad  i=16,18.
\ea
Furthermore, the whole term corresponding to $j=4$ in (\ref{genbox}) is
antisymmetric under $(t\lra u)$.
The following pairs of coefficient functions are negatively related
in the sense of (\ref{bantisym}):
the $b_{i2}^{(5)}$ are related to $b_{l1}^{(5)}$, and
the $b_{i2}^{(6)}$ are related to $b_{l1}^{(6)}$, where $l$ can take any of the
values
$l=i,i+1,i+2$ depending on the value of $i$.
The number of independent coefficient functions is greatly reduced for this
box because of all these relations. We took advantage of this fact when writing
down the relevant coefficient functions in Appendix~A.

As explained after Eq.~(\ref{genbox}) the crossed box (\ref{fig:ggnlot}a4) is
obtained from (\ref{fig:ggnlot}a3) with the help of the
${\cal M}_t \leftrightarrow{\cal M}_u$ operation. For this reason we write
down explicit results only for one of the box (\ref{fig:ggnlot}a3) in Appendix.~A.

A necessary check on the correctness of our one--loop results is gauge invariance.
For example, for gluon 1 this implies that one must have
\be
p_{1\mu} \epsilon_{\nu}(p_2) \bar{u}(p_3)
M^{\mu\nu}({\rm one-loop}) v(p_4)=0,
\ee
for each of the remaining independent amplitude structures that multiply e.g.
$p_{1\nu}$, $p_{3\nu}$ and $\gamma_\nu$. Similarly one must have
\be
p_{2\nu} \epsilon_{\mu}(p_1) \bar{u}(p_3)
M^{\mu\nu}({\rm one-loop}) v(p_4)=0
\ee
again, for each of the remaining independent amplitude structures that multiply e.g.
$p_{2\mu}$, $p_{3\mu}$ and $\gamma_\mu$.
We have verified gauge
invariance for the following gauge-invariant subsets of diagrams: (i) When
the incoming gauge bosons are photons, i.e. including graphs
(\ref{fig:ggnlot}a1), (\ref{fig:ggnlot}c1),
(\ref{fig:ggnlot}c2), (\ref{fig:ggnlot}d1), (\ref{fig:ggnlot}d2),
(\ref{fig:ggnlot}d3) plus their u-channel counterparts
with their
corresponding color weights; (ii) For the photoproduction of
heavy flavors,
i.e. including all the above diagrams plus graphs (\ref{fig:ggnlot}a4),
(\ref{fig:ggnlot}c4), (\ref{fig:ggnlot}e1) plus
their u-channel counterparts, with corresponding color weights;
(iii) For the hadroproduction of heavy flavors, which ultimately includes
all the graphs from Figs.~\ref{fig:ggnlot} and \ref{fig:ggnlos}
plus their relevant u-channel counterparts.
We emphasize that the above gauge invariance checks were made separately
for both color structures $C_F$ and $N_C$, and for every existing
combination of color matrices $T^a$, $T^b$ and $\delta^{ab}$, whenever
they arise. When checking on gauge invariance all the relevant $s$-, $t$- and
$u$-channel graphs have to be added. Gauge invariance must of course be checked
for each power of $\ep$ and for each of the coefficient functions of the
Laurent series expansion of the scalar master integrals separately, independent
of their actual numerical values.


Finally we note that the original computer output for the box diagrams
was extremely long. The final results were cast into the above
shorter form
with the help of the REDUCE Computer Algebra System \cite{reduce}.

\begin{figure*}
\includegraphics{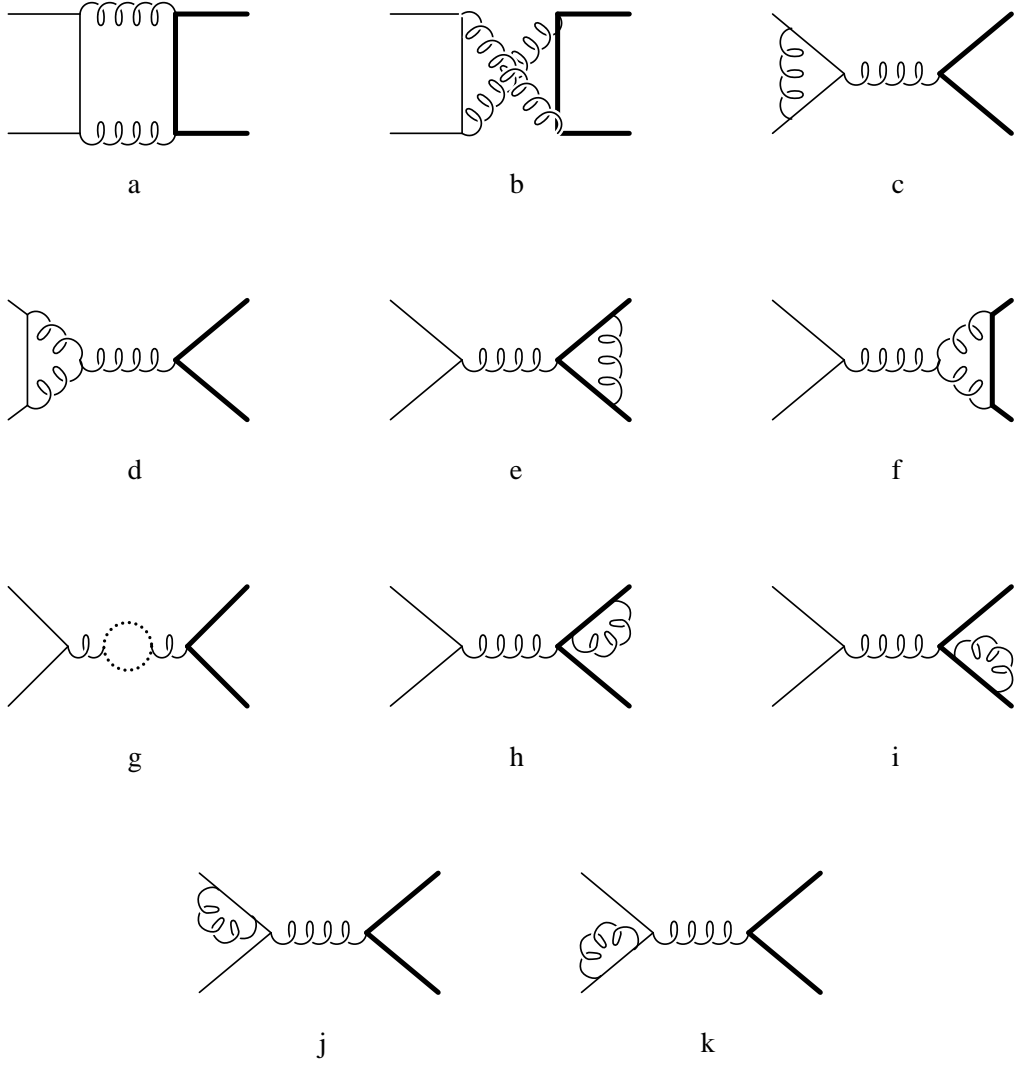}
\caption{\label{fig:qqnlo}
The one-loop Feynman diagrams contributing to the subprocess
$q \bar{q} \rightarrow Q \overline Q$.
The loop with dotted line represents gluon, ghost and light and heavy
quarks.}
\end{figure*}

\section{\label{qqbar}
ANNIHILATION OF THE QUARK-ANTIQUARK PAIR
}

The LO Born graphs contributing to this subprocess are shown in Fig.~\ref{fig:qqlo}.
In Fig.~\ref{fig:qqnlo} we show the graphs contributing at one-loop order.

The leading order contribution proceeds only through the s-channel
graph. One has:
\begin{equation}
\label{qqborn}
B_{q\bar{q}} = i T^a_{ij} T^a_{kl} \bar{v}(p_2)\gamma^{\mu}u(p_1)
                      \bar{u}(p_3)\gamma_{\mu}v(p_4)/s.
\end{equation}
Here the color matrices $T^a$ belong to different fermion lines which are
connected by the gluon having color index $a$.
We have again left out the
factor $g^2$ in the Born term contribution (\ref{qqborn}).
In the Passarino-Veltman reduction for tensor
integrals we can make use of the same scalar integrals of \cite{KMR} as those
appearing in the gluon fusion subprocess, with relevant shifts and interchanges of
momenta when needed.

Starting again with the 2-point insertions, we notice that the result for
graph (6g) can be
obtained from the one of (\ref{grh}) for graph (4h) in the gluon fusion
subprocess by the simple replacement
\begin{equation}
M_{(6\rm g)} = M_{(4\rm h)}^{\mu\nu} \, (B_s^{\mu\nu}
\rightarrow B_{q\bar{q}}).
\end{equation}

The massless quark self-energy insertion graphs (6j) and (6k) with external legs
on-shell vanish identically:
\begin{equation}
M_{(6\rm j)} = M_{(6\rm k)} = 0.
\end{equation}

The massive quark self-energy insertion graphs (6h) and (6i) with external
legs on-shell are calculated in analogy to the ones considered in the
previous section:
\begin{equation}
M_{\rm (6h)} = M_{\rm (6i)} = - C_F B_{q\bar{q}}
\frac{3-2\ep}{\ep (1-2\ep)}\, .
\end{equation}
%

The results for the vertex insertions are relatively short. Starting with graphs
(6c) and (6d) one finds that they are proportional to the LO Born term:
\begin{equation}
M_{(6\rm c)} = B_{q\bar{q}} \sum_{k=1}^2 \ep^k \{
     3 B_5^{(k)} + 2 B_5^{(k-1)} + 2 C_4^{(k)} s   \} /6
\end{equation}
and
\begin{equation}
M_{(6\rm d)} = - 3 B_{q\bar{q}} \sum_{k=1}^2 \ep^k B_5^{(k)}/2.
\end{equation}

For the other two vertex insertion diagrams we also obtain simple expressions:
\begin{eqnarray}
\nonumber
M_{(6\rm e)} &=&  (  B_{q\bar{q}} \sum_{k=1}^2 \ep^k \{
3 B_2^{(k)} + 2 B_2^{(k-1)} + C_6^{(k)} s (1 + \beta^2)    \\
\nn  &&
 - 16 k  \}    + 4 i T^a T^a m
\bar{v}(p_2) {\rm \hspace{-.1in}} \not p_3 u(p_1)    \bar{u}(p_3) v(p_4)
\times     \\
&&  \nn
         \sum_{k=1}^2 \ep^k \{
B_2^{(k)} + 2 B_2^{(k-1)} - 8 k \} /(s^2\beta^2)      )/6       \\
\end{eqnarray}
and
\begin{eqnarray}
\nn
M_{(6\rm f)} &=&  3 ( B_{q\bar{q}} \sum_{k=1}^2 \ep^k \{
   B_5^{(k)} (8 m^2/s - 1) + 2 C_1^{(k)} m^2    \\
\nn    &&
- k 16 (5 m^2/s - 1) \}
        + 4 i T^a T^a m
\bar{v}(p_2) {\rm \hspace{-.1in}} \not p_3 u(p_1) \times    \\
\nn   &&
  \bar{u}(p_3) v(p_4)
       \sum_{k=1}^2 \ep^k \{
B_5^{(k)} (8 m^2/s + 1) - 2 B_5^{(k-1)}    \\
\nn   &&
+ 6 C_1^{(k)} m^2 - C_1^{(k-1)} s
- k 4 (12 m^2/s - 1) \}     \\
&&
/(s^2\beta^2)  )/(2\beta^2).
\end{eqnarray}

Turning to the two box diagrams (6a) and (6b) we note that extensive Dirac algebra
manipulations lead to rather compact expressions for the amplitudes.
We have expanded the box diagrams in terms of seven independent Dirac
structures, the same set for each of the two box graphs.
Then every Dirac structure is multiplied by the sums of products of a
small set of basis functions and coefficient functions.
Thus, we have the following compact expansion for the two box diagrams:
\begin{eqnarray}
\label{boxq}
\nn
M&=&i T_{\rm col} \sum_{k=1}^2 \ep^k \left \{
\bar{v}(p_2) \gamma^{\mu} u(p_1) \bar{u}(p_3) \gamma_{\mu} v(p_4)
                                  \sum f_i^{(k)} h_i^{(0)}  \right. \\
\nonumber  &+&
\bar{v}(p_2) {\rm \hspace{-.1in}} \not p_3 u(p_1)
\bar{u}(p_3) {\rm \hspace{-.1in}} \not p_1 v(p_4)
                           \sum f_i^{(k)} h_{i}^{(1)}   \\
\nonumber  &+&
\bar{v}(p_2)\gamma^{\nu}{\rm \hspace{-.1in}}\not p_3\gamma^{\mu} u(p_1)
\bar{u}(p_3)\gamma_{\mu}{\rm \hspace{-.05in}}\not p_1\gamma_{\nu} v(p_4)
                            \sum f_i^{(k)} h_{i}^{(2)}   \\
\nonumber  &+&
\bar{v}(p_2)\gamma^{\nu}\gamma^{\alpha}\gamma^{\mu} u(p_1)
\bar{u}(p_3)\gamma_{\mu}\gamma_{\alpha}\gamma_{\nu} v(p_4)
                               \sum f_i^{(k)} h_{i}^{(3)}   \\
\nonumber  &+&
m \bar{v}(p_2) {\rm \hspace{-.1in}} \not p_3 u(p_1)
  \bar{u}(p_3) v(p_4)
                                \sum f_i^{(k)} h_{i}^{(4)}   \\
\nonumber  &+&
m \bar{v}(p_2)\gamma^{\mu} u(p_1)
\bar{u}(p_3)\gamma_{\mu}{\rm \hspace{-.05in}}\not p_1 v(p_4)
                                 \sum f_i^{(k)} h_{i}^{(5)}   \\
\nonumber &+&   \left.
m \bar{v}(p_2)\gamma^{\nu}{\rm \hspace{-.1in}}\not p_3\gamma^{\mu} u(p_1)
\bar{u}(p_3)\gamma_{\mu}\gamma_{\nu} v(p_4)
                                    \sum f_i^{(k)} h_{i}^{(6)} \right \}.
\\
\end{eqnarray}

There are seven independent covariants in (\ref{boxq}) upon using the four mass--shell
conditions. We have not attempted to further
reduce the set of seven covariants using Fierz--type identities which
are anyway valid only in $n=4$--dimensions.
Taking parity and the masslessness of the initial quarks into account the number of
amplitudes and thereby the number of independent covariants in $n=4$ is
$2 \cdot 2 \cdot 2 \cdot 2/2 \cdot 2 = 4$. However, this counting may no longer be
true in $n\neq 4$.

The sums over $i$ in (\ref{boxq}) run from 1 to 15 in the box diagram (6a).
Below we list the color factors and analytic functions for the
two 4-point functions of (\ref{boxq}). For the graph (\ref{fig:qqnlo}a) we get:
\begin{equation}
\label{colqa}
T_{\rm col} = (T^a_{im} T^b_{mj}) (T^b_{kn} T^a_{nl}),
\end{equation}
where the first parentheses in (\ref{colqa}) corresponds to the summation
over color indices of the massless fermion line.
The basis functions read
\begin{eqnarray}
\label{fqa}
& f_1^{(k)} = B_1^{(k)}, \qquad  f_2^{(k)} = B_5^{(k)}, &   \\
\nn
& f_3^{(k)} = C_1^{(k-1)},  \qquad    f_4^{(k)} = C_1^{(k)},   \qquad
f_5^{(k)} = C_3^{(k-1)},   &   \\
\nn
& f_6^{(k)} = C_3^{(k)},   \qquad f_7^{(k)} = C_4^{(k-1)},   \qquad
f_8^{(k)} = C_4^{(k)},   &   \\
\nn
& f_9^{(k)} = D_2^{(k-1)},   \qquad      f_{10}^{(k)} = D_2^{(k)},
\qquad      f_{11}^{(k)} = k,   &    \\
\nn
& f_{12}^{(k)} = (k-1) C_1^{(k-2)}, \qquad  f_{13}^{(k)} = (k-1) C_3^{(k-2)},
&  \\   \nn
& f_{14}^{(k)} = (k-1) C_4^{(k-2)}, \qquad  f_{15}^{(k)} = (k-1) D_2^{(k-2)}.&
\end{eqnarray}
As in the case of the gluon fusion boxes there exist a number of universal relations
among the various coefficient functions $h_i^{(j)}$ valid for any value of $j$:
\ba
\label{hrels}
\nn
& h_3^{(j)} = z_t h_7^{(j)}/t, \quad   h_5^{(j)} = 2 t h_7^{(j)}/s,
\quad  h_9^{(j)} = - t h_7^{(j)},   &   \\
\nn
& h_{10}^{(j)} = - t h_8^{(j)},  \quad h_{12}^{(j)} = 2 z_t h_7^{(j)}/t,
\quad  h_{13}^{(j)} = 4 t h_7^{(j)}/s, & \\
& h_{14}^{(j)} = 2 h_7^{(j)}, \quad    h_{15}^{(j)} = - 2 t h_7^{(j)} &
\ea

The color factor for the second box graph
(\ref{fig:qqnlo}b) is
\begin{equation}
\label{colqb}
T_{\rm col} = (T^a T^b) (T^a T^b).
\end{equation}
All basis functions are obtained from those in (\ref{fqa}) by
the interchange $(t\lra u)$, except for the two additional functions
(with subscripts 16 and 17), e.g.:
\begin{eqnarray}
\label{fqb}
& f_1^{(k)} = B_1^{(k)}(t\lra u), \qquad  f_2^{(k)} = B_5^{(k)}, &   \\
\nn
& f_3^{(k)} = C_1^{(k-1)},  \qquad    f_4^{(k)} = C_1^{(k)},  & \\
\nn
& f_5^{(k)} = C_3^{(k-1)}(t\lra u),   \qquad
f_6^{(k)} = C_3^{(k)}(t\lra u),    &   \\
\nn
& \qquad f_7^{(k)} = C_4^{(k-1)},   \qquad
f_8^{(k)} = C_4^{(k)},   &   \\
\nn
& f_9^{(k)} = D_2^{(k-1)}(t\lra u),   \qquad
f_{10}^{(k)} = D_2^{(k)}(t\lra u),     \\
\nn
&     f_{11}^{(k)} = k,   &    \\
\nn
& f_{12}^{(k)} = (k-1) C_1^{(k-2)}, \quad
f_{13}^{(k)} = (k-1) C_3^{(k-2)}(t\lra u),  &  \\
\nn
& f_{14}^{(k)} = (k-1) C_4^{(k-2)}, \quad
f_{15}^{(k)} = (k-1) D_2^{(k-2)}(t\lra u), &   \\
\nn
& f_{16}^{(k)} = B_1^{(k-1)}(t\lra u), \qquad  f_{17}^{(k)} = B_5^{(k-1)}. &
\end{eqnarray}
The last two functions $f_{16}$ and $f_{17}$ appear in the expansion (\ref{boxq})
only in two sums where $h_i^{(1)}$ and $h_i^{(4)}$ are present and, consequently, these
sums run from 1 to 17.

One has further relations for the various coefficient functions $h_i^{(j)}$
which are similar to those in Eq.~(\ref{hrels}). In this case they are valid
for any given value of $j$
except for $j=1$ and $j=4$.
\ba
\label{hqb1rels}
\nn
& h_3^{(j)} = z_u h_7^{(j)}/u, \quad   h_5^{(j)} = 2 u h_7^{(j)}/s,
\quad  h_9^{(j)} = - u h_7^{(j)},   &   \\
\nn
& h_{10}^{(j)} = - u h_8^{(j)},  \quad h_{12}^{(j)} = 2 z_u h_7^{(j)}/u,
\quad  h_{13}^{(j)} = 4 u h_7^{(j)}/s, & \\
& h_{14}^{(j)} = 2 h_7^{(j)}, \quad    h_{15}^{(j)} = - 2 u h_7^{(j)}, &
\ea
where $z_u=2m^2+u$.
In case of $j=1$ and $j=4$ one has
\ba
\label{hqb2rels}
\nn
& h_5^{(j)} = 2 u h_7^{(j)}/s,   \quad  h_9^{(j)} = - u h_7^{(j)},
\quad h_{10}^{(j)} = - u h_8^{(j)},   &  \\
\nn
& h_{12}^{(j)} = z_u h_{14}^{(j)}/u,
\quad  h_{13}^{(j)} = 2 u h_{14}^{(j)}/s, & \\
& h_{15}^{(j)} = - u h_{14}^{(j)}. &
\ea

The coefficient functions $h_i^{(j)}$ are given in Appendix~B of this paper.
However, there exists a partial symmetry for these box diagrams, which
allows one to
express most coefficient functions for the box graph (\ref{fig:qqnlo}b) through
the ones of the
box graph (\ref{fig:qqnlo}a). In particular, starting from the
coefficients $h_i^{(j)}$
with superscript $j\geq 2$, we find the following general relations:
\begin{eqnarray}
\label{relsq}
h_i^{(j)} [{\rm (\ref{fig:qqnlo}b)}] &=& - h_i^{(j)} [{\rm
                                          (\ref{fig:qqnlo}a)}] (t\lra u),
\quad  j=2;                    \\
\nonumber
h_i^{(j)} [{\rm (\ref{fig:qqnlo}b)}] &=& h_i^{(j)} [{\rm
                                          (\ref{fig:qqnlo}a)}] (t\lra u),
\quad  j=3,5,6.
\end{eqnarray}
Consequently, for the graph (\ref{fig:qqnlo}b) only the coefficients
$h_i^{(0)}$, $h_i^{(1)}$ and $h_i^{(4)}$
are presented in Appendix~B. We reiterate that all the one-loop
amplitudes of this chapter must be multiplied by the common factor
(\ref{common}).

\section{\label{summary}
CONCLUSIONS
}

We have presented analytic ${\mathcal O}(\ep^2)$ results on the one-loop
amplitudes
for gluon-- and light quark--induced heavy quark pair production including their
absorptive parts
\footnote{See EPAPS Document No. E-PRVDAQ-73-054605 for our analytical results
          for all the box graphs in REDUCE format.
          For more information on EPAPS, see
          http://www.aip.org/pubservs/epaps.html.}.
These are needed for the calculation of the
loop-by-loop part of the parton model description of NNLO heavy hadron production
in
hadronic collisions. We have not included the finite and divergent pieces in our
presentation since these were already obtained in an earlier publication \cite{KM}.
The advantage of having the results in amplitude form is that one retains the full
spin information of the partonic subprocess which would be of later use when one
wants to consider polarization phenomena in heavy hadron production. As an immediate
next step we plan to square the one--loop amplitudes and to sum over the spins of the
external partons. This
will provide the necessary input for the loop-by-loop part of the NNLO parton model
description of unpolarized heavy hadron or top quark pair production which is
presently under study at the TEVATRON II and will be studied at the upcoming hadron
collider LHC.


\begin{acknowledgments}
Many thanks go to J.~H\"{o}hle for his help in setting up and use
of a REDUCE~3.7 for Linux at the ZDV of the University of Mainz. We are greatful to
J.~Gegelia and
S.~Weinzierl for discussions.
Z.M. is thankful to A.~Pivovarov for giving insight on the current state of
threshold resummations. Z.M. would like to thank the Particle Theory group of the
Institut f{\"u}r Physik, Universit{\"a}t Mainz, for hospitality.
The work of Z.M. was supported by a DFG (Germany) grant under contract
436 GEO 17/4/04 and
partly by the Graduiertenkolleg ``Eichtheorien'' at the University of Mainz.
M.R. was supported by the DFG through the Graduiertenkolleg ``Eichtheorien''
at the University of Mainz.
\end{acknowledgments}

\appendix
\section{}

Here we present the coefficients of the box contributions for the gluon
fusion subprocess appearing in Eq.~(\ref{genbox}).

We define a shorthand notation:
\begin{eqnarray}
\nn  &&
z_1 \equiv m^2 s - t^2,   \qquad    z_2 \equiv s + 2 t,     \\  &&
z_t \equiv 2 m^2 + t,     \qquad    z_u \equiv 2 m^2 + u,    \\
\nn   &&
D \equiv m^2 s - u t.
\end{eqnarray}

First we list coefficients for the abelian type of box diagram
(\ref{fig:ggnlot}a1):
\ba
\nn   &&
b_{i,1}^{(0)} = 0, \qquad  i=1,2,3,4,13,     \\
\nn   &&
b_{5,1}^{(0)}= - 10 t z_t/D,  \qquad    b_{6,1}^{(0)}= 2 t z_u/D,   \\
\nn   &&
b_{8,1}^{(0)}= s z_u/D,      \qquad     b_{9,1}^{(0)}= - 5 s t \beta^2/D,  \\
\nn   &&
b_{10,1}^{(0)}= - s u_1 \beta^2/D,    \qquad    b_{12,1}^{(0)}= s (D + m^2 s \beta^2)/D,    \\
\nn   &&
b_{14,1}^{(0)}= - 24 t z_t/D;     \\   \nn
\\   \nn  &&
b_{1,1}^{(1)}=0,        \qquad            b_{2,1}^{(1)}= - 2 z_t^2/t D,\qquad b_{3,1}^{(1)}=0,     \\
\nn   &&
b_{4,1}^{(1)}= 2 z_t/D, \qquad  b_{5,1}^{(1)}=2 t z_t ( 6 D + s t \beta^2 )/D^2,  \\
\nn &&
b_{6,1}^{(1)}=- 2 z_t ( 2 m^2 D - s t^2 \beta^2 )/D^2 , \\
\nn &&
b_{8,1}^{(1)}=( 2 m^2 z_2 D + s^2 t z_t \beta^2 )/D^2, \\
\nn &&
b_{9,1}^{(1)}=s \beta^2/(2 z_t)  b_{5,1}^{(1)}  ,
\qquad
b_{10,1}^{(1)}=s^2 t^2 \beta^4/D^2 , \\
\nn &&
b_{13,1}^{(1)}=16 m^2 z_t/t D , \qquad
b_{14,1}^{(1)}=4 t z_t ( 6 D + s t \beta^2 )/D^2  ;\\ \nn
\\ \nn &&
b_{1,2}^{(1)}=4 T u \beta^2/D^2,\\
\nn &&
b_{2,2}^{(1)}=- 4 m^2 ( 2 D^2 + 2 t^2 D
                             + s t (D-2 t^2) \beta^2 )/ s t^2 D^2 ,\\
\nn &&
b_{3,2}^{(1)}= 4 t u z_t/sD^2,\qquad
b_{4,2}^{(1)}= 4 m^2 ( D - 2 s z_t )/s D^2,\\
\nn &&
b_{5,2}^{(1)}= 4 t z_t ( 2 m^2 D - t (D+m^2 s) \beta^2 )/D^3,\\
\nn &&
b_{6,2}^{(1)}= - 8 m^2 ( T (s T-2 m^2 t) D + s t^3 z_t \beta^2 )/t D^3,\\
\nn &&
b_{8,2}^{(1)}= 4 m^2 t (u z_2 D - s^2 z_1 \beta^2)/s D^3,\\
\nn &&
b_{9,2}^{(1)}= 2 t^2 \beta^2 ( (2 m^2-s) D - m^2 s^2 \beta^2 )/D^3, \\
\nn &&
b_{10,2}^{(1)}= 4 m^2 t \beta^2 ( (2 s+t) D - s^2 t \beta^2 )/D^3,\\
\nn &&
b_{13,2}^{(1)}= 16 m^2 ( (2 s T + t z_t) D - 3 m^2 s t z_t )/ s t^2 D^2, \\
\nn &&
b_{14,2}^{(1)}= 4 t z_t ( 4 m^2 D - t (D+3 m^2 s) \beta^2 )/D^3; \\
\nn\\
\nn &&
b_{1,3}^{(1)}=-4 ( 2 D^2 - 2 m^2 (3 m^2-s) D + m^4 s t \beta^2 )/s T D^2,\\
\nn &&
b_{2,3}^{(1)}= - 4 z_t ( 2 m^2 D^2 - t (2 m^4 D - (3 s+2 t) z_t D \\
\nn && \shift  - 2 m^2 t^2 z_u) )/s t^2 T D^2,\\
\nn &&
b_{3,3}^{(1)}= 4 ( z_t D - t (m^2 s-u (2 m^2-s)) \beta^2 )/s \beta^2 D^2,\\
\nn &&
b_{4,3}^{(1)}= 4 ( s^2 (m^2-t) \beta^4 + (m^2+4 s) \beta^2 D \\
\nn && \shift - t z_1 \beta^2
                                                   + z_t D )/s \beta^2 D^2,\\
\nn &&
b_{5,3}^{(1)}= - 4 t ( 2 (2 m^2+2 s-t) D^2 - t (6 m^2 z_2+s^2 \beta^2) D\\
\nn && \shift         + m^2 s (6 s D+2 t D+s t z_u) \beta^2 )/sD^3, \\
\nn &&
b_{6,3}^{(1)}= - 4 ( 4 m^2 T D^2 - t ( (2 m^4-t^2) z_t
                       - 2 m^2 t z_2 ) D\\
\nn && \shift + 2 m^4 s t^2 z_2 \beta^2 )/tD^3,\\
\nn &&
b_{8,3}^{(1)}= - 2 ( 2 m^2 s D^2 - (2 m^2 u^3-s^2 t^2 \beta^2
                           -2 m^2 t^3) D \\
\nn && \shift- 2 m^2 s^2 z_1 u \beta^2 )/sD^3,\\
\nn &&
b_{9,3}^{(1)}= - 2 t ( D^2 + (9 m^2 s-2 m^2 t+s t-3 t^2) D \beta^2\\
\nn && \shift     + s^2 (3 D-m^2 u) \beta^4 )/D^3, \\
\nn &&
b_{10,3}^{(1)}= - 2 z_t ( t z_t D + s^2 t^2 \beta^4\\
\nn && \shift                          - (2 m^2 s^2 z_2 - m^2 s t u + t^4) \beta^2 )/D^3,\\
\nn &&
b_{13,3}^{(1)}= 16 m^2 ( s (4 D + t (8 m^2+5 s)) D \beta^2\\
\nn && \shift
                                - 3 t^2 (D+m^2 s \beta^2) z_2
                                  + 3 m^2 s^3 t \beta^4 )/s^2 t^2 \beta^2 D^2,\\
\nn &&
b_{14,3}^{(1)}= 4 t z_t ( 2 (2 s T-3 s z_2-3 t u) D\\
\nn && \shift   + s u (D+3 m^2 s) \beta^2 )/sD^3; \\
\nn\\
\nn &&
b_{1,4}^{(1)}= 4 T z_t z_2/sD^2,\\
\nn &&
b_{2,4}^{(1)}= 8 T ( z_t D + s t^2 \beta^2 )/s t D^2,\\
\nn &&
b_{3,4}^{(1)}= - 4 ( s z_t D + s t (D+t z_t) \beta^2 )/s^2 \beta^2 D^2,\\
\nn &&
b_{4,4}^{(1)}= - 4 ( 3 T \beta^2 D + z_t D - 2 t^2 z_t \beta^2 )/s \beta^2 D^2,\\
\nn &&
b_{5,4}^{(1)}= - 4 t z_t ( 6 D^2 - 4 t T D + s^2 t T \beta^2 )/sD^3,\\
\nn &&
b_{6,4}^{(1)}= 4 z_t ( (2 m^4-t^2) D - 2 s t^2 T \beta^2 )/D^3,\\
\nn &&
b_{8,4}^{(1)}= - 2 ( 2 m^2 (3 s+2 t) D^2 + 3 s^2 t T D \beta^2\\
\nn && \shift
                            + m^2 s t z_2 D - 2 s^2 t^3 z_t \beta^2 )/sD^3,\\
\nn &&
b_{9,4}^{(1)}= 2 t ( 4 m^2 D^2 - s (6 m^2 s-6 m^2 t+7 s t+2 t^2) D
                                               \beta^2 \\
\nn && \shift+ s^2 t^3 \beta^4 )/sD^3, \\
\nn &&
b_{10,4}^{(1)}= 2 t ( D^2 - (3 m^2 s-10 m^2 t+4 s t+t^2) D \beta^2\\
\nn && \shift
                                                     + 2 s t^3 \beta^4 )/D^3,\\
\nn &&
b_{13,4}^{(1)}= 16 m^2 z_t ( 6 t D - s (2 s T-t^2) \beta^2 )/s^2 t \beta^2 D^2,\\
\nn &&
b_{14,4}^{(1)}= - 4 t z_t ( 2 (6 m^2 s-8 m^2 t+7 s t+3 t^2) D\\
\nn && \shift
                                                   - 3 s t^3 \beta^2 )/s D^3;\\
\nn\\
\nn &&
b_{i 5}^{(1)}= b_{i 2}^{(1)} ;  \\
                                           \\
\nn &&
b_{1,1}^{(2)}= 4 T u/sD,\\
\nn &&
b_{2,1}^{(2)}= - 2 T ( 2 D + s t )/s t D,\\
\nn &&
b_{3,1}^{(2)}= - 4 u ( D - t z_t )/s^2 \beta^2 D,\\
\nn &&
b_{4,1}^{(2)}= - 2 ( (4 m^2-3 s) D + s t z_t )/s^2 \beta^2 D,\\
\nn &&
b_{5,1}^{(2)}= 2 t^2 z_t ( 4 m^2 s+t z_2 )/sD^2,\\
\nn &&
b_{6,1}^{(2)}= - 2 ( 2 (m^4-t T) D + t^3 z_t )/D^2,\\
\nn &&
b_{8,1}^{(2)}= ( 2 (m^2 t+s z_t) D - s t^2 z_t )/D^2,\\
\nn &&
b_{9,1}^{(2)}= t ( 4 m^2 u D + s t (4 m^2 s+t z_2) \beta^2 )/sD^2,\\
\nn &&
b_{10,1}^{(2)}= t ( 4 m^2 u D + s^2 (2 D-t^2) \beta^2 )/sD^2,\\
\nn &&
b_{13,1}^{(2)}= 16 m^2 ( s^2 T \beta^2 + 3 t u z_t )/s^2 t \beta^2 D,\\
\nn &&
b_{14,1}^{(2)}= 4 t^2 z_t ( 4 m^2 s + 2 s t + 3 t^2 )/sD^2;\\
\nn \\
\nn &&
b_{1,2}^{(2)}= - 4 ( D - m^2 u )/sD,\\
\nn &&
b_{2,2}^{(2)}= - 2 ( 2 (s+T) D + m^2 s t )/s t D,\\
\nn &&
b_{3,2}^{(2)}= - 4 t ( m^2 z_2 + s z_u )/s^2 \beta^2 D,\\
\nn &&
b_{4,2}^{(2)}= 2 ( D \beta^2 - t z_u )/s \beta^2 D,\\
\nn &&
b_{5,2}^{(2)}= - 2 t ( 2 z_t (4 s-t) D + m^2 s^2 t \beta^2\\
\nn && \shift
                                      + 3 m^2 s t z_2 - s t^2 z_u )/sD^2,\\
\nn &&
b_{6,2}^{(2)}= - 2 ( 2 T^2 D + t^3 z_u )/D^2,\\
\nn &&
b_{8,2}^{(2)}= - ( 2 D^2 - 2 m^2 t D + t^2 u z_2 )/D^2,\\
\nn &&
b_{9,2}^{(2)}= - t ( t D + (6 s D-5 m^2 s t+3 s^2 t-3 t^3) \beta^2 )/D^2,\\
\nn &&
b_{10,2}^{(2)}= - t^2 ( 4 m^2 D - s^2 u \beta^2 )/sD^2,\\
\nn &&
b_{13,2}^{(2)}= 16 ( s ((m^2+s) D+2 m^2 s t) \beta^2\\
\nn && \shift
                                    - m^2 t^2 (2 T-u) )/s^2 t \beta^2 D,\\
\nn &&
b_{14,2}^{(2)}= - 4 t z_t ( 10 s D - 2 m^2 s z_2 + 3 t^2 u )/sD^2;\\
\nn\\
\nn &&
b_{1,1}^{(3)}= 4 ( m^4 s + T D )/s t D,\\
\nn &&
b_{2,1}^{(3)}= 2 ( 2 (s+T) D + m^2 s (4 m^2+t) )/s t D,\\
\nn &&
b_{3,1}^{(3)}= - 4 ( s D + m^2 s^2 \beta^2 - m^2 t z_2 )/s^2 \beta^2 D,\\
\nn &&
b_{4,1}^{(3)}= - 2 ( 2 s D + 3 m^2 s^2 \beta^2 - 2 m^2 t z_2 )/s^2 \beta^2 D,\\
\nn &&
b_{5,1}^{(3)}= 2 ( 4 m^2 D^2 + 4 t (6 m^2 s+m^2 t+s t) D\\
\nn && \shift
                              + s t^3 (5 s+3 t) \beta^2 + 2 t^3 z_1 )/sD^2,\\
\nn &&
b_{6,1}^{(3)}= 2 ( (2 m^2 T+t^2) D - m^2 s t^2 \beta^2 + 2 m^4 t z_2 )/D^2,\\
\nn &&
b_{7,1}^{(3)}= ( (24 m^2 s-8 m^2 t+7 s t) D \\
\nn && \shift
- s^2 t (3 m^2-2 t) \beta^2
                                                        + 2 t^2 z_1 )/D^2,\\
\nn &&
b_{8,1}^{(3)}= ( 2 m^2 (s-t) D - 2 m^2 s^2 t \beta^2 - s t^2 z_u )/D^2,\\
\nn &&
b_{9,1}^{(3)}= - t ( u D - ((8 s-3 t) D+m^2 s (5 s+6 t)) \beta^2 )/D^2,\\
\nn &&
b_{10,1}^{(3)}= ( 2 (s^2 T+2 m^2 t^2) D
                                     - s^2 (2 m^2 z_1-t^2 u) \beta^2 )/sD^2,\\
\nn &&
b_{13,1}^{(3)}= 16 ( (16 m^4-s^2) D + m^2 t u (5 z_t - 4 z_2) )
                                                          /s^2 t \beta^2 D,\\
\nn &&
b_{14,1}^{(3)}= 4 t z_t ( (10 s-3 t) D + m^2 s (4 s+5 t) )/sD^2;\\
\nn \\
\nn &&
b_{1,2}^{(3)}= 4 T ( m^2 s + D )/s t D,\\
\nn &&
b_{2,2}^{(3)}= 2 T ( 3 s z_t + 2 t^2 )/s t D,\\
\nn &&
b_{3,2}^{(3)}= - 4 ( s D \beta^2 + m^2 t z_2 )/s^2 \beta^2 D,\\
\nn &&
b_{4,2}^{(3)}= - 2 ( (3 s T+t^2) \beta^2 + t z_t )/s \beta^2 D,\\
\nn &&
b_{5,2}^{(3)}= 2 ( 4 m^2 D^2 - 2 t^3 D + s t^4 \beta^2
                                                  + 4 m^2 t^2 z_1 )/sD^2,\\
\nn &&
b_{6,2}^{(3)}= 2 ( 2 m^2 T D - 2 s t^2 T \beta^2 - t^3 z_t )/D^2,\\
\nn &&
b_{7,2}^{(3)}= - t ( 2 t D - s t^2 \beta^2 - 4 m^2 z_1 )/D^2,\\
\nn &&
b_{8,2}^{(3)}= - t ( 2 s^2 T \beta^2 + 2 m^2 D + s t z_t )/D^2,\\
\nn &&
b_{9,2}^{(3)}= t^2 ( 4 m^2 D + s (4 m^2 s-t z_2) \beta^2 )/sD^2,\\
\nn &&
b_{10,2}^{(3)}= - ( 2 (4 m^4 s+2 m^2 t u+s^2 t) D\\
\nn && \shift
                                      + s^2 (4 m^2 s T-t^3) \beta^2 )/sD^2,\\
\nn &&
b_{13,2}^{(3)}= - 16 m^2 ( 4 s^2 T \beta^2 - 3 t^2 z_t )/s^2 t \beta^2 D,\\
\nn &&
b_{14,2}^{(3)}= 4 t^2 z_t ( 2 m^2 s - 2 s t - 3 t^2 )/sD^2;\\
\nn \\
\nn &&
b_{1,1}^{(4)}=0, \qquad               b_{2,1}^{(4)}=0,\\
\nn &&
b_{3,1}^{(4)}= 4/s \beta^2, \qquad      b_{4,1}^{(4)}= 4/s \beta^2,\\
\nn &&
b_{5,1}^{(4)}= 2 t^2/D,  \qquad     b_{6,1}^{(4)}= 2 t (s + 3 t)/D,\\
\nn &&
b_{9,1}^{(4)}= - t z_2/D, \qquad   b_{10,1}^{(4)}= - ( 2 D + s^2 \beta^2 + t z_2 )/D,\\
\nn &&
b_{13,1}^{(4)}= - 24/s \beta^2, \qquad   b_{14,1}^{(4)}=0;\\
\nn \\
\nn &&
b_{1,1}^{(5)}= 4 T/tD,  \qquad      b_{2,1}^{(5)}= 4 T/tD,\\
\nn &&
b_{3,1}^{(5)}= 4 z_t/s D \beta^2, \qquad  b_{4,1}^{(5)}= 4 z_t/s D \beta^2,\\
\nn &&
b_{5,1}^{(5)}= 4 ( T D + 2 t^2 z_t )/D^2, \qquad b_{6,1}^{(5)}= 4 t T z_2/D^2,  \\
\nn &&
  b_{7,1}^{(5)}= 2 t ( D + 2 z_1 )/D^2,\\
\nn &&
b_{9,1}^{(5)}= - 2 t ( D - 2 s t \beta^2 )/D^2,\\
\nn &&
b_{10,1}^{(5)}= - 2 T ( 2 D + s^2 \beta^2 )/D^2,\\
\nn &&
b_{13,1}^{(5)}= - 24 ( m^2 s \beta^2 - t z_u )/s t D \beta^2,\\
\nn &&
b_{14,1}^{(5)}= 12 t^2 z_t/D^2;\\
\nn \\
\nn &&
b_{1,2}^{(5)}= 4 m^2/tD, \qquad       b_{2,2}^{(5)}= 4 m^2/tD,\\
\nn &&
b_{3,2}^{(5)}= 4 z_u/s D \beta^2,\qquad    b_{4,2}^{(5)}= 4 z_u/s D \beta^2,\\
\nn &&
b_{5,2}^{(5)}= 4 ( T D + 2 m^2 t z_2 )/D^2, \qquad b_{6,2}^{(5)}= 4 t^2 z_u/D^2,\\
\nn &&
b_{7,2}^{(5)}= 2 ( 2 m^2 s z_2 - t D )/D^2,\\
\nn &&
b_{9,2}^{(5)}= 2 t ( D - 2 s u \beta^2 )/D^2,\\
\nn &&
b_{10,2}^{(5)}= 2 ( (s \beta^2+2 T) D - m^2 s^2 \beta^2 )/D^2,\\
\nn &&
b_{13,2}^{(5)}= - 24 ( m^2 s \beta^2 + t z_u )/s t D \beta^2,\\
\nn &&
b_{14,2}^{(5)}= - 12 t u z_t/D^2;\\
\nn \\
\nn &&
b_{1,1}^{(7)}=0,\quad                b_{2,1}^{(7)}= - 2 z_t/D,\\
\nn &&
b_{3,1}^{(7)}=0, \quad               b_{4,1}^{(7)}= - 2 ( 4 D - s t \beta^2 )/s D \beta^2,\\
\nn &&
b_{5,1}^{(7)}= - 2 t z_t ( 4 D + t z_2 )/D^2,\\
\nn &&
b_{6,1}^{(7)}= - 2 t ( 6 T D - s t^2 \beta^2 )/D^2,\\
\nn &&
b_{9,1}^{(7)}= s \beta^2/(2 z_t) b_{5,1}^{(7)},\\
\nn &&
b_{10,1}^{(7)}= t ( 8 m^2 D + 2 s t u \beta^2 + t z_2^2 )/D^2,\\
\nn &&
b_{13,1}{(7)}= 16 ( m^2 s \beta^2 + 2 D )/s D \beta^2,\\
\nn &&
b_{14,1}^{(7)}=  - 4 t z_t ( 4 D + t z_2 )/D^2;\\
\nn\\
\nn &&
b_{1,2}^{(7)}= - 4 ( (2 m^2 s-t z_2) D - m^2 t^2 z_2 )/s t D^2,\\
\nn &&
b_{2,2}^{(7)}= - 4 ( 4 D^2 - s t D + 2 m^2 t^2 z_2 )/s t D^2,\\
\nn &&
b_{3,2}^{(7)}= 4 ( 2 T D + (2 s D+t^2 u) \beta^2 )/s \beta^2 D^2,\\
\nn &&
b_{4,2}^{(7)}= - 4 ( (3 m^2 z_2-2 s^2-3 s t) D\\
\nn && \shift                                       + 2 m^2 s^2 t \beta^2 )/s^2 \beta^2 D^2,\\
\nn &&
b_{5,2}^{(7)}= - 4 ( 2 D^3 - 4 t (2 m^2-s) D^2 \\
\nn && \shift+ 2 t^2 u (2 s+3 t) D
                                               + m^2 s^2 t^3 \beta^2 )/sD^3,\\
\nn &&
b_{6,2}^{(7)}= 4 t ( (6 m^4-2 m^2 s+4 m^2 t+t^2) D\\
\nn && \shift                                                 - 2 m^2 s t^2 \beta^2 )/D^3,\\
\nn &&
b_{7,2}^{(7)}= 2 ( 4 z_t D^2 + 2 u (2 m^2 s-t^2) D\\
\nn && \shift                                                 - m^2 s^2 t^2 \beta^2 )/D^3,\\
\nn &&
b_{9,2}^{(7)}= 2 t ( t (2 s T - 12 m^2 t - s^2) D\\
 \nn && \shift                               + s^2 (4 (m^2-s) D+t^2 z_u) \beta^2 )/sD^3,\\
\nn &&
b_{10,2}^{(7)}= - 2 ( 12 m^2 t D^2 - s^2 (2 m^2 s+t^2) D \beta^2\\
 \nn && \shift                                   - 2 m^2 t^2 z_2 (D+s^2 \beta^2) )/sD^3,\\
\nn &&
b_{13,2}^{(7)}= 16 ( t (2 m^2 u-s z_2) D      \\
 \nn && \shift               + s (5 m^2 s D+t^2 D-3 m^2 t^2 u) \beta^2 )
                                                        /s^2 t \beta^2 D^2,\\
\nn &&
b_{14,2}^{(7)}= 4 t z_t ( 2 s (2 z_t-3 s) D - 3 t^2 (s^2 \beta^2-t z_2) )/sD^3;  \\
\nn \\
\nn &&
b_{1,3}^{(7)}= - 4 ( 2 z_t u D - m^4 t z_2 )/s T D^2,\\
\nn &&
b_{2,3}^{(7)}= - 4 ( 4 T D^2 - t (4 m^2 u-s t) D  \\
 \nn && \shift                                     + 2 m^4 t^2 z_2 )/s t T D^2, \\
\nn &&
b_{3,3}^{(7)}= 4 ( (2 T+s) D + (m^2 s^2+t^2 u) \beta^2 )/s \beta^2 D^2,\\
\nn &&
b_{4,3}^{(7)}= - 4 ( (m^2 s+6 m^2 t+s^2) D\\
\nn && \shift                                   + 2 s^2 t U \beta^2 )/s^2 \beta^2 D^2,\\
\nn &&
b_{5,3}^{(7)}= 4 t ( 2 (4 m^2-u) D^2 \\
\nn && \shift - t (s^2-2 t^2) D+ m^2 s^2 t u \beta^2 )/sD^3,\\
\nn &&
b_{6,3}^{(7)}= 4 t ( (6 m^4-2 m^2 t-t^2) D + 2 m^2 s t u\beta^2 )/D^3,\\
\nn &&
b_{9,3}^{(7)}= - 2 t ( 4 D^2 - t z_t D\\
 \nn && \shift                    - ((4 m^2 s-3 s^2+3 t^2) D-m^2 s u z_2) \beta^2 )/D^3,\\
\nn &&
b_{10,3}^{(7)}= 2 ( 8 m^2 u D^2 + m^2 s t z_2 D \\
\nn && \shift + s^2 t (m^2-t) D \beta^2
                                         - 2 m^2 s^2 t u z_2 \beta^2 )/sD^3,\\
\nn &&
b_{13,3}^{(7)}= 16 ( 2 t u (3 m^2-s) D \\
\nn && \shift + m^2 (4 s D+3 t u^2) s \beta^2 )
                                                        /s^2 t \beta^2 D^2,\\
\nn &&
b_{14,3}^{(7)}= 4 t z_t ( (8 m^2 s-2 s t+6 t^2) D\\
\nn && \shift
                                     - 3 s^3 t \beta^2 + 3 s t^2 z_t )/sD^3;\\
\nn \\
\nn &&
b_{1,4}^{(7)}= - 4 T ( 2 s D - t^2 z_2 )/s t D^2, \\
\nn &&
b_{2,4}^{(7)}= - 8 T ( s D + t^2 z_2 )/s t D^2, \\
\nn &&
b_{3,4}^{(7)}= - 4 ( 2 (3 m^2+2 t) D + t^3 \beta^2 )/s \beta^2 D^2, \\
\nn &&
b_{4,4}^{(7)}= - 4 ( m^2 (s-2 t) D - 2 s t^3 \beta^2 )/s^2 D^2 \beta^2, \\
\nn &&
b_{5,4}^{(7)}= - 4 ( 2 D^3 - 8 m^2 t D^2 \\
\nn && \shift- 2 t^3 (2 s+3 t) D+ s^2 t^3 T \beta^2 )/sD^3,  \\
\nn &&
b_{6,4}^{(7)}= 4 t ( (6 m^2 T-2 s T+t^2) D - 2 s t^2 T \beta^2 )/D^3, \\
\nn &&
b_{7,4}^{(7)}= 2 ( 12 m^2 D^2 - (4 m^2 s T-3 t^2 z_2) D
                                                       + s t^4 \beta^2 )/D^3, \\
\nn &&
b_{9,4}^{(7)}= 2 t ( 3 D^2 - 2 m^2 z_t D\\
  \nn && \shift                             + (3 (m^2 s+t^2) D+s t^2 z_t) \beta^2 )/D^3, \\
\nn &&
b_{10,4}^{(7)}= 2 ( 2 (2 m^2 u+s z_2) D^2 - 2 m^2 t^2 z_2 D\\
   \nn && \shift                                  + s^2 t z_t (D+2 t^2) \beta^2 )/sD^3, \\
\nn &&
b_{13,4}^{(7)}= 16 ( 2 m^2 t (2 s+3 t) D  \\
\nn && \shift                        + 3 s (D^2+m^2 t^3) \beta^2 )/s^2 t \beta^2 D^2, \\
\nn &&
b_{14,4}^{(7)}= 4 t z_t ( 2 (4 m^2 s+2 s t+3 t^2) D + 3 s t^2 z_t )/sD^3; \\
\nn\\
\nn &&
b_{1,5}^{(7)}= - 4 T u z_2/sD^2, \\
\nn &&
b_{2,5}^{(7)}= - 4 ( 2 D^2 + t (D+2 m^2 t) z_2 )/s t D^2, \\
\nn &&
b_{3,5}^{(7)}= - 4 ( 2 m^2 D - t^2 u \beta^2 )/s \beta^2 D^2, \\
\nn &&
b_{4,5}^{(7)}= 4 ( (s T-6 m^2 t) D - 2 m^2 s^2 t \beta^2 )/s^2 D^2 \beta^2, \\
\nn &&
b_{5,5}^{(7)}= 4 t z_t ( 4 D^2 + m^2 s t z_2 )/sD^3, \\
\nn &&
b_{6,5}^{(7)}= - 4 t ( 2 D^2 - (6 m^2 T+t z_t) D
                                                 + 2 m^2 s t^2 \beta^2 )/D^3, \\
\nn &&
b_{9,5}^{(7)}= - 2 t ( 4 (2 m^2-s) D^2 + 2 m^2 t z_2 D\\
  \nn && \shift                                    - m^2 s^2 (2 D+t z_2) \beta^2 )/sD^3, \\
\nn &&
b_{10,5}^{(7)}= - 2 ( 2 (2 m^2-s) z_2 D^2 + 2 m^2 s t z_t D\\
  \nn && \shift                                  - s^2 t^2 (D+2 m^2 z_2) \beta^2 )/sD^3, \\
\nn &&
b_{13,5}^{(7)}= 16 ( 2 m^2 t^2 D + s (2 D^2-3 m^2 t^2 u) \beta^2 )
                                                        /s^2 t \beta^2 D^2, \\
\nn &&
b_{14,5}^{(7)}= 4 t z_t ( 8 D^2 - t z_2 D + 3 m^2 s t z_2 )/sD^3.
\ea
%
%

%
%
Next we list the coefficients for the nonabelian box diagram (\ref{fig:ggnlot}a2):
\ba
\nn   &&
b_{1,1}^{(0)}= 0,  \qquad  b_{2,1}^{(0)}= 2z_t/t^2,  \qquad  b_{3,1}^{(0)}= 0,\\
\nn   &&
  b_{4,1}^{(0)}= 0, \qquad b_{6,1}^{(0)}= - s(4T + s)/D,  \\
\nn   &&
          b_{7,1}^{(0)}= - 16t^2/D, \qquad b_{8,1}^{(0)}= - 2( sz_t + 4t^2 )/D,\\
\nn   &&
b_{10,1}^{(0)}= s( s - 2t )/D, \qquad    b_{12,1}^{(0)}= s( sz_t + 4t^2 )/D, \\
\nn   &&
b_{13,1}^{(0)}= - 16m^2/t^2; \\
\nn\\
\nn   &&
b_{1,1}^{(1)}=0,      \qquad               b_{2,1}^{(1)}= 2z_t/D, \\
\nn   &&
b_{3,1}^{(1)}=0,        \qquad             b_{4,1}^{(1)}= 2(m^2s + t^2)/sD, \\
\nn   &&
b_{6,1}^{(1)}= ( 2(sT-4m^2t)D - t^2z_2^2 )/D^2, \\
\nn   &&
b_{7,1}^{(1)}= 2t^2( 8D - sz_t )/D^2, \\
\nn   &&
  b_{8,1}^{(1)}= 2t( 4TD + t^2z_2 )/D^2, \\
\nn   &&
b_{10,1}^{(1)}= ( D^2 - 4t^2D + st^2z_2 )/D^2, \\
\nn   &&
b_{12,1}^{(1)}= t^3( 4D - sz_2 )/D^2, \qquad b_{13,1}^{(1)}= - 16m^2/D; \\
\nn \\
\nn   &&
b_{1,2}^{(1)}= - 4Tuz_t/tD^2,\\
\nn   &&
 b_{2,2}^{(1)}= - 4m^2( D + 2tz_t )/tD^2,\qquad b_{3,2}^{(1)}= - 4t^2u/sD^2, \\
\nn   &&
b_{4,2}^{(1)}= 4m^2( - D + 2st\beta^2 )/s\beta^2D^2, \\
\nn   &&
b_{5,2}^{(1)}= 2( (4m^4s-2m^2t^2+st^2)D + m^2s^2t^2\beta^2 )/D^3, \\
\nn   &&
b_{6,2}^{(1)}= 4m^2( - 2m^2D^2 - s^2tD\beta^2 + s^3t^2\beta^4 )
                                                           /s\beta^2D^3, \\
\nn   &&
b_{7,2}^{(1)}= 4t^2( 2m^2sT - t^2u )/D^3, \qquad b_{8,2}^{(1)}= 8m^2t^2z_1/D^3, \\
\nn   &&
b_{13,2}^{(1)}= 16m^2( D + 3m^2s\beta^2 )/s\beta^2D^2, \\
\nn   &&
b_{15,2}^{(1)}= 4t^2( TD + 3m^2z_1 )/D^3; \\
\nn\\
\nn   &&
b_{1,3}^{(1)}= 4m^2z_t( 2D + m^2t )/tTD^2, \\
\nn   &&
b_{2,3}^{(1)}= - 4( (2T(5m^2s+4m^2t-t^2)+3t^2z_2)D\\
\nn   && \shift        - 2m^2t^3z_u )/stTD^2, \\
\nn   &&
b_{3,3}^{(1)}= 4tu^2/sD^2, \\
\nn   &&
b_{4,3}^{(1)}= 4( (u(8m^2-3s)+sT)D - 2m^2s^2u\beta^2 )
                    /s^2\beta^2D^2, \\
\nn   &&
b_{5,3}^{(1)}= 2( 4tD^2 + ((10m^2s+t^2)(2m^2-s)-2s^2t)D \\
 \nn   && \shift             - st^2u^2\beta^2 )/D^3, \\
\nn   &&
b_{6,3}^{(1)}= 2( 2m^2(3z_t+s+t)D^2 - s^2(2m^2s-t^2)D\beta^2\\
     \nn   && \shift                        - 2m^4s^4\beta^4 )/s\beta^2D^3, \\
\nn   &&
b_{7,3}^{(1)}= 4t^2( (3s(2m^2-s)+2tz_2)D\\
   \nn   && \shift      - m^2s^3\beta^2 + m^2stz_2 )/sD^3, \\
\nn   &&
b_{8,3}^{(1)}= 4tz_t( tD - 2m^2su )/D^3,\\
\nn   &&
b_{13,3}^{(1)}= -16m^2( 4(t(5m^2+2t)-s^2\beta^2)D\\
    \nn   && \shift                         + s(2m^2su-s^2z_t+t^3)\beta^2 )
                                                      /s^2t\beta^2D^2, \\
\nn   &&
b_{15,3}^{(1)}= 4t^2( 8D^2 - s(7s+5t)D\\
 \nn   && \shift                       + 3m^2s(-s^2\beta^2+tz_2) )/sD^3; \\
\nn\\
\nn   &&
b_{1,4}^{(1)}= 4Tz_t/D^2, \qquad
b_{2,4}^{(1)}= 8tTz_2/sD^2, \\
\nn   &&
b_{3,4}^{(1)}= 4t^3/sD^2, \\
\nn   &&
b_{4,4}^{(1)}= - 4( (3m^2+2t)D + 2t^3\beta^2 )/s\beta^2D^2, \\
\nn   &&
b_{5,4}^{(1)}= - 2( 4(3m^2+2t)D^2\\
 \nn   && \shift+ t^2(2m^2-s-4t)D   + st^4\beta^2 )/D^3, \\
\nn   &&
b_{6,4}^{(1)}= 2( 2(2m^4-sz_t)D^2 - s^2(2m^2T+tz_t)D\beta^2\\
          \nn   && \shift    - 2st^2(m^2z_1-t^2z_t)\beta^2 )/s\beta^2D^3, \\
\nn   &&
b_{7,4}^{(1)}= - 4t^2( 6D^2 + stD - t^3z_2 )/sD^3, \\
\nn   &&
b_{8,4}^{(1)}= - 4t^2( (2T+t)D + 2t^2z_t )/D^3, \\
\nn   &&
b_{13,4}^{(1)}= 16m^2( 2(2T+s+2t)D - 3st^2\beta^2 )/s^2\beta^2D^2, \\
\nn   &&
b_{15,4}^{(1)}= - 4t^2( 12D^2 - 2tuD - 3t^3z_2 )/sD^3; \\
\nn\\
\nn\\
\nn &&
b_{i 5}^{(1)}= b_{i 2}^{(1)} ;  \\
                                           \\
\nn   &&
b_{1,1}^{(2)}=0, \qquad b_{2,1}^{(2)}= - 2T( 3s + 4t )/sD, \\
\nn   &&
b_{3,1}^{(2)}=0, \qquad
b_{4,1}^{(2)}= - ( (D-2t^2)\beta^2 - 4m^2z_t )/s\beta^2D, \\
\nn   &&
b_{6,1}^{(2)}= ( - 2(2m^2-s)D^2 + 2s(m^2s+tT)D\beta^2\\
               \nn   && \shift - tz_t(4m^2D-s^2(4m^2+t)\beta^2) )/s\beta^2D^2, \\
\nn   &&
b_{7,1}^{(2)}= 2t^3( 8m^2s + 5st + 4t^2 )/sD^2, \\
\nn   &&
b_{8,1}^{(2)}= - 2t( 2sTz_t + t^3 )/D^2, \\
\nn   &&
b_{10,1}^{(2)}= - ( 2m^2s^2T - 4t^2D + t^3z_2 )/D^2, \\
\nn   &&
b_{12,1}^{(2)}= - t^2( 2sD + t(8m^2s + 5st + 4t^2) )/D^2, \\
\nn   &&
b_{13,1}^{(2)}= - 16( sTu\beta^2 + 2m^4z_2 )/s^2\beta^2D; \\
\nn\\
\nn   &&
b_{1,2}^{(2)}=0,\\
\nn   &&
b_{2,2}^{(2)}= 2( 2(m^2s-2t^2)D - m^2t^2(3s+4t) )/st^2D, \\
\nn   &&
b_{3,2}^{(2)}=0,\\
\nn   &&
b_{4,2}^{(2)}= ( tz_2 - (3m^2s-2st-t^2)\beta^2 )/s\beta^2D, \\
\nn   &&
b_{6,2}^{(2)}= ( 2m^2tz_2D - s(sz_t-4tT)D\beta^2\\
  \nn   && \shift     - m^2s^2tz_2\beta^2 - 2m^2s^3t\beta^4 )/s\beta^2D^2, \\
\nn   &&
b_{7,2}^{(2)}= - 2t^2( 8(s-t)D - tu(3s+4t) )/sD^2, \\
\nn   &&
b_{8,2}^{(2)}= - 2( D^2 + 5t^2D + m^2st(4m^2+t) )/D^2, \\
\nn   &&
b_{10,2}^{(2)}= - ( 2m^2s^2T - 4t^2D - t^2u(3s+2t) )/D^2, \\
\nn   &&
b_{12,2}^{(2)}= - t^2( 2(s+4t)D + tu(3s+4t) )/D^2, \\
\nn   &&
b_{13,2}^{(2)}= 8( - 2sD^2\beta^2 + st(2m^2t(3s+2t)-3uD)\beta^2\\
     \nn   && \shift                          - 2m^2t^3z_2 )/s^2t^2\beta^2D; \\
\nn\\
\nn   &&
b_{1,1}^{(3)}=0, \\
\nn   &&
b_{2,1}^{(3)}= - 2( 2(m^2s-2tz_t)D + m^2t^2(5s+4t) )/st^2D, \\
\nn   &&
b_{3,1}^{(3)}=0, \\
\nn   &&
b_{4,1}^{(3)}= - ( 8m^2D + s(7m^2s+3t^2)\beta^2 + 3s^2z_t )
                                                             /s^2\beta^2D, \\
\nn   &&
b_{6,1}^{(3)}= - ( D^2 + 3m^2z_2D - (10m^2s-4m^2t\\
\nn   && \shift
+2st-3t^2)D\beta^2
      + m^2st(2s\beta^2+3z_2)\beta^2 )/\beta^2D^2, \\
\nn   &&
b_{7,1}^{(3)}= 2t^2( 16sD + tu(5s+4t) )/sD^2, \\
\nn   &&
b_{8,1}^{(3)}= 2( D^2 + 5t^2D + m^2t^2(5s+4t) )/D^2, \\
\nn   &&
b_{10,1}^{(3)}= - ( 2D^2 - 10stD - 4m^2st^2 - 5st^2u )/D^2, \\
\nn   &&
b_{12,1}^{(3)}= t( 4D^2 - 5stD - m^2st(5s+4t) )/D^2, \\
\nn   &&
b_{13,1}^{(3)}= 8( t(16m^4+su)D
+ m^2s^2(2D-t(4m^2-t))\beta^2\\
\nn   && \shift                            + m^2t^2(5s+4t)z_2 )/s^2t^2\beta^2D; \\
\nn\\
\nn   &&
b_{1,2}^{(3)}=0, \qquad
b_{2,2}^{(3)}= 2T( 8m^2s + 3st + 4t^2 )/stD, \\
\nn   &&
b_{3,2}^{(3)}=0, \\
\nn   &&
b_{4,2}^{(3)}= ( - (D+s(4m^2+t))\beta^2 + 3sz_t )/s\beta^2D, \\
\nn   &&
b_{6,2}^{(3)}= ( 12m^2D^2 + 2t(m^2-s)z_2D\\
    \nn   && \shift  + st(4m^2z_1-stz_t)\beta^2 )/s\beta^2D^2, \\
\nn   &&
b_{7,2}^{(3)}= - 2t^4( 5s+4t )/sD^2, \\
\nn   &&
b_{8,2}^{(3)}= 2t^2( 2D + 4m^2t - t^2 )/D^2, \\
\nn   &&
b(_{10,2}^{(3)}= - ( 2m^2s^2T + 3st^3 + 2t^4 )/D^2, \\
\nn   &&
b_{12,2}^{(3)}= t( 2sz_tD + t^3(5s+4t) )/D^2, \\
\nn   &&
b_{13,2}^{(3)}= - 16( s((4m^2+t)D+m^2t^2)\beta^2 \\
 \nn   && \shift + 6m^4tz_2 )
                                                          /s^2t\beta^2D; \\
\nn\\
\nn   &&
b_{1,1}^{(4)}= 0,  \qquad            b_{2,1}^{(4)}= 2/t, \qquad b_{3,1}^{(4)}= 0,\\
\nn   &&
 b_{4,1}^{(4)}= 2/s\beta^2, \qquad b_{6,1}^{(4)}= ( D + sz_2\beta^2 )/\beta^2D,\\
\nn   &&
b_{7,1}^{(4)}= 0, \qquad b_{8,1}^{(4)}= - 2st/D,  \\
\nn   &&
b_{13,1}^{(4)}= 8(4m^2 + u)/st\beta^2; \\
\nn\\
\nn   &&
b_{1,1}^{(5)}= 0, \qquad             b_{2,1}^{(5)}= 8T/tD,\qquad b_{3,1}^{(5)}= 0, \\
\nn   &&
b_{4,1}^{(5)}= 8z_t/s\beta^2D,\qquad b_{6,1}^{(5)}= 2( z_2D + 2stz_t\beta^2 )
                                                                 /\beta^2D^2,  \\
\nn   &&
b_{7,1}^{(5)}= 8t^3/D^2, \qquad b_{8,1}^{(5)}= - 4t( D - 2t^2 )/D^2, \\
\nn   &&
b_{13,1}^{(5)}= 32( - m^2s\beta^2 + tz_u )/st\beta^2D; \\
\nn\\
\nn   &&
b_{1,2}^{(5)}= 0,   \qquad        b_{2,2}^{(5)}= 8m^2/tD,\qquad b_{3,2}^{(5)}= 0, \\
\nn   &&
  b_{4,2}^{(5)}= 8z_u/s\beta^2D,\\
\nn   &&
 b_{6,2}^{(5)}= - 2z_2( D - 2m^2s\beta^2 )/\beta^2D^2, \\
\nn   &&
b_{7,2}^{(5)}= - 8t^2u/D^2, \qquad b_{8,2}^{(5)}= - 4t(D + 2tu)/D^2, \\
\nn   &&
b_{13,2}^{(5)}= - 32( m^2s\beta^2 + tz_u )/st\beta^2D; \\
\nn\\
\nn   &&
b_{1,1}^{(7)}= 0,    \qquad          b_{2,1}^{(7)}= 2t/D,\qquad b_{3,1}^{(7)}= 0, \\
\nn   &&
 b_{4,1}^{(7)}= - 2z_t/\beta^2D, \\
\nn   &&
b_{6,1}^{(7)}= - ( 4\beta^2D^2 + sz_tD + stz_1\beta^2 )/\beta^2D^2, \\
\nn   &&
b_{7,1}^{(7)}= - 2t^2( 4D + st )/D^2, \\
\nn   &&
b_{8,1}^{(7)}= - 2st^3/D^2, \qquad b_{13,1}^{(7)}= 16m^2z_2/s\beta^2D; \\
\nn\\
\nn   &&
b_{1,2}^{(7)}= - 4Tu/D^2, \\
\nn   &&
b_{2,2}^{(7)}= - 4( 6D^2 + t(z_t+3u)D + 2t^3u )/stD^2, \\
\nn   &&
b_{3,2}^{(7)}= 4( z_tD + m^2(2D+tz_2)\beta^2 )/s\beta^4D^2,   \\
\nn   &&
b_{4,2}^{(7)}= - 4( 3m^2z_2D + (4m^2t-3sz_t-4s^2\beta^2)D\beta^2\\
\nn && \shift                                  + 2m^2stz_2\beta^2 )/s^2\beta^4D^2, \\
\nn   &&
b_{5,2}^{(7)}= 2( z_tD^2 + sTz_tD\beta^2+(8m^4s-8m^2s^2\\
 \nn && \shift     +4m^2t^2-4s^2t-3st^2+t^3)D\beta^4\\
   \nn && \shift                            + st(2m^4s+t^2u)\beta^4 )/\beta^4D^3, \\
\nn   &&
b_{6,2}^{(7)}= 2( - 12m^4z_2D^2
                  + s^2(2m^2z_1\\
 \nn && \shift -5m^2sz_2-st^2)D\beta^4   + m^2s(4tD+s^2z_2)D\beta^2 \\
  \nn && \shift          + 2m^2s^3tz_1\beta^4 )/s^2\beta^4D^3, \\
\nn   &&
b_{7,2}^{(7)}= 4t^2( 4D^2 + 2u(2s+t)D + m^2s^2t )/sD^3, \\
\nn   &&
b_{8,2}^{(7)}= 4t( (4m^2s+2m^2t+t^2)D + 2t^3u )/D^3, \\
\nn   &&
b_{13,2}^{(7)}= 16( 16m^4tz_tD + 2m^2s^3(2D+m^2s)\beta^4\\
 \nn && \shift
                     - s^2t(4m^4s+7m^2tz_u+2tu^2)\beta^2 )
                                                    /s^3t\beta^4D^2, \\
\nn   &&
b_{15,2}^{(7)}= 4t^2( 2(-s^2\beta^2+u(3s-2t))D + 3st^2u )/sD^3; \\
\nn\\
\nn   &&
b_{1,3}^{(7)}= 4m^2(2D + m^2t)/TD^2, \\
\nn   &&
b_{2,3}^{(7)}= - 4( 4TD^2 + t(2m^2T-3st-2t^2)D \\
 \nn && \shift
                                          + 2m^2t^3u )/stTD^2, \\
\nn   &&
b_{3,3}^{(7)}= 4z_u( 4m^2D + s(2D-tu)\beta^2 )/s^2\beta^4D^2, \\
\nn   &&
b_{4,3}^{(7)}= 4( 2m^2(4m^2z_2+sz_t)D - 3s^2(m^2-s)D\beta^2   \\
   \nn && \shift               - 2s^2(2m^2t^2+tu(s+3t))\beta^2 )/s^3\beta^4D^2, \\
\nn   &&
b_{5,3}^{(7)}= 2( - 2D^3 + 2tz_tD^2
               - 2m^2(3s+2t)D^2\beta^2 \\
\nn && \shift+ 2s(4m^2-3s)D^2\beta^4
               - st(s^2+4m^2t)D\beta^4\\
        \nn && \shift        + 2m^2st(6m^2-t)D\beta^2
               - m^2s^2tz_2^2\beta^2\\
           \nn && \shift     - m^2s^2t^2(3s+2t)\beta^4 )/s\beta^4D^3, \\
\nn   &&
b_{6,3}^{(7)}= 2( 3m^2z_2D^2 + m^2(8m^2+3s+18t)D^2\beta^2\\
\nn && \shift  + s^2t(10m^2+t)D\beta^2
               + 2m^2s^2(z_t+2u)D\beta^4\\
          \nn && \shift     - 2m^2s^2tz_2^2\beta^2
          - 2m^2s^2t^2(3s+2t)\beta^4 )/s\beta^4D^3, \\
\nn   &&
b_{7,3}^{(7)}= 4t^2( 2(2m^2s-s^2+t^2)D + s^2tU )/sD^3, \\
\nn   &&
b_{8,3}^{(7)}= 4t( (4m^2s+tz_t)D - 2m^2stu )/D^3, \\
\nn   &&
b_{13,3}^{(7)}= 16( 2t(8m^4z_t+s^2z_u)D + 10m^2s^2tD\beta^2  \\
       \nn && \shift    + m^2s^3(4D+3st)\beta^4 - 3m^2s^2t^2z_t\beta^2 )
                                                       /s^3t\beta^4D^2, \\
\nn   &&
b_{15,3}^{(7)}= 4t^2( 2(-s^2\beta^2+u(s-2t))D - 3stu^2 )/sD^3; \\
\nn\\
\nn   &&
b_{1,4}^{(7)}= 4tT/D^2, \\
\nn   &&
      b_{2,4}^{(7)}= - 8T( (2s+t)D + st^2 )/stD^2, \\
\nn   &&
b_{3,4}^{(7)}= - 4z_t( 4m^2D - st^2\beta^2 )/s^2\beta^4D^2, \\
\nn   &&
b_{4,4}^{(7)}= b_{4,3}^{(7)} - 4 (3sD + 2tz_2^2)\beta^2 )/s\beta^2D^2, \\
\nn   &&
b_{5,4}^{(7)}= b_{5,3}^{(7)} - 2 ( 4z_uD^2 - 2s^2(m^2+2t)D\beta^2 \\
\nn && \shift                              + s^3t^2\beta^4 )
                                                             /\beta^2D^3,  \\
\nn   &&
b_{6,4}^{(7)}= b_{6,3}^{(7)} - 4( 2(2m^2z_2+sT)D^2  \\
\nn && \shift   - s^2(m^2s-3t^2)D\beta^2 + s^4t^2\beta^4 )/s\beta^2D^3, \\
\nn   &&
b_{7,4}^{(7)}= 4t^2( 2(2m^2s+t^2)D + s^2tT )/sD^3, \\
\nn   &&
b_{8,4}^{(7)}= 4t( (2sT+tz_t)D - 2t^4 )/D^3, \\
\nn   &&
b_{13,4}^{(7)}= b_{13,3}^{(7)} - 16 ( (4m^2-3s)D + 3 t z_2 z_u )
                                                        /s\beta^2D^2, \\
\nn   &&
b_{15,4}^{(7)}= 4t^2( 2(4m^2s+tz_2)D - 3st^3 )/sD^3; \\
\nn\\
\nn   &&
b_{1,5}^{(7)}= - 4Tu/D^2, \\
\nn   &&
b_{2,5}^{(7)}= - 4( 2D^2 + t(2T+s)D + 2t^3u )/stD^2, \\
\nn   &&
b_{3,5}^{(7)}= 4( z_tD - m^2(sz_t-4D)\beta^2 )/s\beta^4D^2, \\
\nn   &&
b_{4,5}^{(7)}= 4( 3(2m^2t-sT)D  - 2m^2z_2D\beta^2\\
   \nn && \shift                      - 2st^2z_u\beta^2 )/s^2\beta^4D^2, \\
\nn   &&
b_{5,5}^{(7)}= 2( 2(16m^6-8m^4t+s^2t)D^2\\
   \nn && \shift            + 2m^2s(3sD-2m^2tz_2)D\beta^2\\
  \nn && \shift             + 2m^2s^3(3m^2+4t)D\beta^4
               - m^2st^2z_2^3\beta^2 )/s^2\beta^4D^3, \\
\nn   &&
b_{6,5}^{(7)}= 2( 2(5m^2z_t-sz_2)D^2
               + 2m^2(6s+t)D^2\beta^2\\
      \nn && \shift + 4m^2t^2z_2D\beta^2
               + st(4m^2u-st)D\beta^4\\
     \nn && \shift + 2m^2s(D^2 - st^2z_2)\beta^4 )/s\beta^4D^3, \\
\nn   &&
b_{7,5}^{(7)}= 4t^2( 4D^2 - tz_2D + st^2u )/sD^3,   \\
\nn   &&
b_{8,5}^{(7)}= 4t( 2D^2 + tz_tD + 2t^3u )/D^3, \\
\nn   &&
b_{13,5}^{(7)}= - 16( tz_uD + m^2t(tz_t+sz_2)\beta^2\\
      \nn && \shift    + u(2TD+t^2u)\beta^4 )/st\beta^4D^2, \\
\nn   &&
b_{15,5}^{(7)}= 4t^2( 8D^2 - 2tz_2D + 3st^2u )/sD^3.
\ea
%
%
%

%
%
Finally, the coefficients for the crossed box (\ref{fig:ggnlot}a4) are:
\ba
\nn   &&
b_{1,1}^{(0)}= 0, \qquad   b_{2,1}^{(0)}= -z_t/t^2,  \qquad  b_{3,1}^{(0)}= 0, \\
\nn   &&
b_{5,1}^{(0)}= 7 t (2 D + t u)/sD  ,     \\
\nn   &&
b_{6,1}^{(0)}= t (2 D + s^2 - t u)/sD ,\\
\nn   &&
b_{10,1}^{(0)}= ( 2 (t-u) D - t (s^2-t u) )/sD, \\
\nn   &&
b_{14,1}^{(0)}= ( 2 (t^2+u^2) D + t u (s^2-t u) )/sD ,\\
\nn   &&
b_{15,1}^{(0)}= 8 m^2 (t^2 + u^2)/t^2 u^2; \\
\nn\\
\nn   &&
b_{1,1}^{(1)}=0     ,\qquad                b_{2,1}^{(1)}= 2 m^2 z_t/tD ,     \\
\nn   &&
b_{3,1}^{(1)}=0      ,\qquad               b_{4,1}^{(1)}= 2 z_u (D - m^2 u)/u^2D ,\\
\nn   &&
b_{5,1}^{(1)}= t (2 D + t u) ( t z_u - 6 D )/sD^2 ,\\
\nn   &&
b_{6,1}^{(1)}= ( 2 (2 m^2-t) D^2 + 2 s t^2 D - t^2 u^2 z_t )/sD^2 ,\\
\nn   &&
b_{8,1}^{(1)}= - ( 2 z_u D^2 - 2 s t u D + t u^3 z_t )/sD^2 ,\\
\nn   &&
b_{10,1}^{(1)}= - t ( 4 D^2 - 2 t^2 D + t^2 u z_u )/sD^2 ,\\
\nn   &&
b_{12,1}^{(1)}= t ( 4 D^2 + 2 t u D - t u^2 z_u )/sD^2 ,\\
\nn   &&
b_{15,1}^{(1)}= 16 m^2 ( - t D + m^2 u (t-u) )/t u^2 D ;\\
\nn\\
\nn   &&
b_{1,2}^{(1)}= 4 T u z_u/sD^2 ,\\
\nn   &&
b_{2,2}^{(1)}= 4 m^2 ( 2 D^2 + t (s-2 T) D - 2 t^2 u z_t )
                                                           /s t^2 D^2 ,\\
\nn   &&
b_{3,2}^{(1)}= 4 t^2 U z_u/s u D^2 ,\\
\nn   &&
b_{4,2}^{(1)}= 4 m^2 ( (z_t-u) D + 2 m^2 u (t-u) )/s u D^2 ,\\
\nn   &&
b_{5,2}^{(1)}= 2 t ( 2 m^2 (2 m^2 t^2-2 m^2 u^2-t^2 u) D\\
\nn && \shift
                                                + t^2 u^3 z_t )/s^2D^3 ,\\
\nn   &&
b_{6,2}^{(1)}= - 4 m^2 ( 2 (2 t+u) D^3 + t^3 u (2 t-u) D\\
\nn && \shift                            + t^4 u^2 (t-u) )/s^2 t D^3 ,\\
\nn   &&
b_{8,2}^{(1)}= 4 m^2 ( 2 D^3 - t u^2 (2 t-u) D
                                            - t^2 u^3 (t-u) )/s^2D^3 ,\\
\nn   &&
b_{9,2}^{(1)}= - 2 t^2 u^2 ( 2 m^2 D - t^2 z_u )/s^2D^3 ,\\
\nn   &&
b_{10,2}^{(1)}= 4 m^2 t^2 u ( t D + m^2 s (t-u) )/s^2D^3 ,\\
\nn   &&
b_{15,2}^{(1)}= - 16 m^2 ( 2 (m^2 t^2-u T (2 t+u)) D \\
 \nn && \shift                - 3 t^2 u^2 z_t )    /s t^2 u D^2 ,\\
\nn   &&
b_{16,2}^{(1)}= - 2 t (2 D + t u) ( 4 m^4 s (t-u) + t u^2 z_t )
                                                                 /s^2D^3 ;\\
\nn\\
\nn   &&
b_{1,3}^{(1)}= - 4 ( m^2 t^2 D + u (m^4 u-t^2 U) z_t )/s t T D^2 ,\\
\nn   &&
b_{2,3}^{(1)}= 4 ( 2 T (t+T) D^2 + t (2 m^4 s-m^4 z_t+t^2 u) D\\
  \nn && \shift                                   + 2 m^2 t T u^2 z_t )/s t^2 T D^2 ,\\
\nn   &&
b_{3,3}^{(1)}= - 4 t U z_u/sD^2 ,\\
\nn   &&
b_{4,3}^{(1)}= 4 U ( z_u D - 2 u^2 z_t )/s u D^2 ,\\
\nn   &&
b_{5,3}^{(1)}= 2 t ( - 12 D^3 + 4 u (2 m^2-t) D^2\\
  \nn && \shift          - u^2 (5 u z_t+s t) D - t u^4 z_t )/s^2D^3 ,\\
\nn   &&
b_{6,3}^{(1)}= 2 ( - 2 (4 m^2 t+2 m^2 u+3 t^2) D^3 - 2 t^3 u D^2\\
 \nn && \shift  + t^2 u^2 (t z_t-4 m^2 u) D
                                 + 2 m^2 t^3 u^3 (t-u) )/s^2 t D^3 ,\\
\nn   &&
b_{8,3}^{(1)}= 2 ( 2 (2 m^2-3 u) D^3 - 2 t u^2 D^2\\
 \nn && \shift            + u^3 (t z_t-4 m^2 u) D
                                        + 2 m^2 t u^4 (t-u) )/s^2D^3 ,\\
\nn   &&
b_{9,3}^{(1)}= 2 t^2 u ( 6 D^2 - u (t+2 U) D + u^3 z_t )/s^2D^3 ,\\
\nn   &&
b_{10,3}^{(1)}= 2 t u ( 2 t D^2 - u (t z_t-4 m^2 u) D\\
  \nn && \shift - 2 m^2 t u^2 (t-u) )/s^2D^3 ,\\
\nn   &&
b_{15,3}^{(1)}= - 16 m^2 ( 4 u D^2 + 2 m^2 (t^2+u^2) D\\
 \nn && \shift              + 3 t u^3 z_t )/s t^2 u D^2 ,\\
\nn   &&
b_{16,3}^{(1)}= - 2 t (2 D + t u) ( 12 D^2 - 2 u (z_t+2 U) D\\
   \nn && \shift + 3 u^3 z_t )/s^2D^3 ;\\
\nn\\
\nn   &&
b_{1,1}^{(2)}= - 4 T u/sD ,\qquad
b_{2,1}^{(2)}= 2 T (2 D - t u)/s t D ,\\
\nn   &&
b_{3,1}^{(2)}= 4 (m^2 t - D)/sD ,\\
\nn   &&
b_{4,1}^{(2)}= - 2 U ( 2 D + t^2 )/s u D ,\\
\nn   &&
b_{5,1}^{(2)}= - t^2 ( 2 D^2 + 10 m^2 s D + 3 t^2 u^2 )/s^2D^2 ,\\
\nn   &&
b_{6,1}^{(2)}= ( (s T+D) (2 m^4 s^2-t^2 u^2) - 2 t^4 D )/s^2D^2 ,\\
\nn   &&
b_{8,1}^{(2)}= ( -2 (2 m^2 s+t u) D^2 \\
\nn   && \shift + 2 s u (2 m^2 u-s t) D
                                                    - t^3 u^3 )/s^2D^2  ,\\
\nn   &&
b_{9,1}^{(2)}= t^3 u ( 8 D + 3 t u )/s^2D^2 ,\\
\nn   &&
b_{10,1}^{(2)}= t ( 4 s D^2 + 2 t (s^2-2 u^2) D + t^3 u^2 )/s^2D^2 ,\\
\nn   &&
b_{12,1}^{(2)}= ( -2 (m^2 s^2 (t-u)-2 t^2 u^2) D + t^3 u^3 )/s^2D^2 ,\\
\nn   &&
b_{14,1}^{(2)}= t^2 ( 2 (2 m^2 s^2+t^2 u+u^3) D - t^2 u^3 )/s^2D^2 ,\\
\nn   &&
b_{15,1}^{(2)}= - 8 ( 4 m^2 t^2 u - (2 m^2 (t-u)+s t) D )/s t u D ,\\
\nn   &&
b_{16,1}^{(2)}= - 4 t^2 ( 2 D + m^2 s ) ( D + m^2 s )/s^2D^2 ;\\
               \\
\nn   &&
b_{1,2}^{(2)}= 4 ( D - m^2 u )/sD ,\\
\nn   &&
b_{2,2}^{(2)}= 2 ( 2 (m^2 u+2 t T) D + t^3 U )/s t^2 D ,\\
\nn   &&
b_{3,2}^{(2)}= 4 t U/sD ,\\
\nn   &&
b_{4,2}^{(2)}= 2 ( (t z_u-2 u^2) D - m^2 u^3 )/s u^2 D ,\\
\nn   &&
b_{5,2}^{(2)}= - t ( 2 (8 m^2 s u+7 s t z_u+t u^2) D
                                                  - 3 t^2 u^3 )/s^2D^2 ,\\
\nn   &&
b_{6,2}^{(2)}= ( 2 (2 m^4 s^2+4 m^2 s^2 t+m^2 s t u-s t^2 u-t^4) D\\
\nn   && \shift
                                                    + t^3 u^3 )/s^2D^2 ,\\
\nn   &&
b_{8,2}^{(2)}= ( -2 (2 m^4 s^2+2 m^2 s t u+s T u^2+t^3 u) D\\
\nn   && \shift                                                    + t^2 u^4 )/s^2D^2 ,\\
\nn   &&
b_{9,2}^{(2)}= t^2 u ( 2 (7 t+3 u) D - 3 t u^2 )/s^2D^2 ,\\
\nn   &&
b_{10,2}^{(2)}= (2 s (t-u) D^2 - 2 t^3 (s-2 u) D - t^3 u^3)/s^2D^2 ,\\
\nn   &&
b_{12,2}^{(2)}= - t ( 2 (2 m^2 s^2+t^2 u-t u^2) D + t u^4 )/s^2D^2 ,\\
\nn   &&
b_{15,2}^{(2)}= 8 ( (2 m^2 s u^2-2 m^2 t (t^2+u^2)-s t u^2) D\\
  \nn   && \shift                        + 4 m^2 t^2 u^3 )/s t^2 u^2 D ,\\
\nn   &&
b_{16,2}^{(2)}= - 4 t (2 D+t u) ( (7 t+4 u) D - t u^2 )/s^2D^2 ;\\
\nn\\
\nn   &&
b_{1,1}^{(3)}= 0 ,\\
\nn   &&
b_{2,1}^{(3)}= - 2 ( (8 m^2+t-2 u) D + 5 m^2 t u )/s t D ,\\
\nn   &&
b_{3,1}^{(3)}= 0 ,\\
\nn   &&
b_{4,1}^{(3)}= - 2 ( (t z_u-2 u (3 m^2+u)) D + 5 m^2 u^3 )/s u^2 D ,\\
\nn   &&
b_{5,1}^{(3)}= t (2 D+t u) ( 2 (7 t+8 u) D + 5 t u^2 )/s^2D^2  ,\\
\nn   &&
b_{6,1}^{(3)}= t ( 2 (s t z_t+3 m^2 s u+2 t (t^2+u^2)) D \\
\nn   && \shift  + 5 t^2 u^3 )  /s^2D^2 ,\\
\nn   &&
b_{10,1}^{(3)}= t^2 ( 2 (4 m^2 s+5 s u-t^2) D - 5 t u^3 )/s^2D^2 ,\\
\nn   &&
b_{12,1}^{(3)}= - ( 4 (t^2+u^2) D^2 - 2 t u (s t-5 u^2) D\\
\nn   && \shift                                               + 5 t^2 u^4 )/s^2D^2  ,\\
\nn   &&
b_{15,1}^{(3)}= - 16 ( (s (m^2 t-u^2)+4 m^2 u (t-u)) D\\
  \nn   && \shift                                          - 5 m^2 t u^3 )/s t u^2 D ;\\
\nn\\
\nn   &&
b_{1,2}^{(3)}= 0 ,\qquad
b_{2,2}^{(3)}= - 2 T ( 8 D + 5 t u )/s t D ,\\
\nn   &&
b_{3,2}^{(3)}= 0 ,\\
\nn   &&
b_{4,2}^{(3)}= - 2 ( (2 m^2 t-6 m^2 u+3 t u) D - 5 m^2 t u^2 )
                                                              /s u^2 D ,\\
\nn   &&
b_{5,2}^{(3)}= - t^2 (2 D + t u) (2 D + 5 t u)/s^2D^2 ,\\
\nn   &&
b_{6,2}^{(3)}= - t^2 ( 2 (m^2 s-s^2+4 t u) D + 5 t^2 u^2 )/s^2D^2 ,\\
\nn   &&
b_{10,2}^{(3)}= t^2 ( 2 (4 m^2 s-s^2+t u) D + 5 t^2 u^2 )/s^2D^2 ,\\
\nn   &&
b_{12,2}^{(3)}= - ( 2 (2 m^2 s (t^2+u^2)-s^2 t u-t^2 u^2) D\\
  \nn   && \shift                                                 - 5 t^3 u^3 )/s^2D^2 ,\\
\nn   &&
b_{15,2}^{(3)}= - 16 ( (4 m^2 u (t-u)+s t U) D \\
 \nn   && \shift+ 5 m^2 t^2 u^2 )                /s t u^2 D ;\\
\nn\\
\nn   &&
b_{i,1}^{(4)}= 0,\qquad         i=1,3,5,9,16,\\
\nn   &&
       b_{2,1}^{(4)}= - 1/t ,\qquad       b_{4,1}^{(4)}= 1/u ,\qquad     b_{6,1}^{(4)}= t (u - t)/D ,\\
 \nn   &&          b_{10,1}^{(4)}= t (t - u)/D ,\qquad b_{15,1}^{(4)}= - 4 (t - u)/tu  ;\\
\nn\\
\nn   &&
b_{1,1}^{(5)}= - 4 T/tD ,\qquad        b_{2,1}^{(5)}= - 4 T/tD ,\\
\nn   &&
b_{3,1}^{(5)}= 4 m^2/uD ,\qquad        b_{4,1}^{(5)}= 4 m^2/uD ,\\
\nn   &&
b_{5,1}^{(5)}= - 4 ( D^2 + m^2 s t^2 )/sD^2 ,\qquad b_{6,1}^{(5)}= 2 t T u/D^2 ,\\
\nn   &&
b_{7,1}^{(5)}= 4 ( D^2 - m^2 s t u )/sD^2 ,\qquad b_{9,1}^{(5)}= 4 t^3 u/sD^2 ,\\
\nn   &&
b_{10,1}^{(5)}= - 2 t T u/D^2 ,\qquad b_{15,1}^{(5)}= - 24 ( 2 m^2 t + D )/t u D ,\\
\nn   &&
b_{16,1}^{(5)}= - 6 t^2 ( 2 D + t u )/sD^2 ;\\
\nn\\
\nn   &&
b_{1,1}^{(6)}= 0 ,\qquad        b_{2,1}^{(6)}= - 8 m^2/tD , \qquad b_{3,1}^{(6)}= 0,\\
\nn   &&
       b_{4,1}^{(6)}= 8 U/uD ,    \qquad  b_{5,1}^{(6)}= 4 t u ( 2 D + t u )/sD^2,  \\
\nn   &&
b_{6,1}^{(6)}= 2 t ( 2 m^2 u + D )/D^2 ,\qquad b_{10,1}^{(6)}= - b_{6,1}^{(6)} ,\\
\nn   &&
b_{15,1}^{(6)}= 32 ( 2 m^2 u + D )/t u D ;\\
\nn\\
\nn   &&
b_{1,1}^{(7)}=0  ,\qquad              b_{2,1}^{(7)}= 2 m^2/D ,\qquad b_{3,1}^{(7)}=0,\\
\nn   &&
           b_{4,1}^{(7)}= 2 (m^2 u - D)/uD ,\\
\nn   &&
b_{5,1}^{(7)}= t ( 4 D - t u ) ( 2 D + t u )/sD^2 ,\\
\nn   &&
b_{6,1}^{(7)}= t ( 6 D^2 - 2 t^2 D - t^2 u^2 )/sD^2 ,\\
\nn   &&
b_{10,1}^{(7)}= t^3 ( 2 D + u^2 )/sD^2 ,\\
\nn   &&
b_{15,1}^{(7)}= 8 ( D - 2 m^2 u )/uD ;\\
\nn\\
\nn   &&
b_{1,2}^{(7)}= - 4 T u^2/s D^2 ,\\
\nn   &&
b_{2,2}^{(7)}= 4 ( (5 m^2 s-T u) D + 2 t T u^2 )/s t D^2 ,\\
\nn   &&
b_{3,2}^{(7)}= - 4 t^2 U/sD^2 ,\\
\nn   &&
b_{4,2}^{(7)}= - 4 ( (2 m^2 s-T u-4 u^2) D + 2 m^2 t u^2 )/s u D^2 ,\\
\nn   &&
b_{5,2}^{(7)}= 2 t ( - 8 D^3 + 4 u (t+2 u) D^2\\
 \nn   && \shift + t u^2 (7 t+4 u) D + t^3 u^3 )/s^2D^3 ,\\
\nn   &&
b_{6,2}^{(7)}= 2 t ( - 6 D^3 + u^2 (4 m^2 s+5 t^2) D\\
 \nn   && \shift + 2 t^3 u^3 )/s^2D^3 ,\\
\nn   &&
b_{9,2}^{(7)}= 2 t^2 u ( 4 D^2 - u (3 t+4 u) D - t^2 u^2 )/s^2D^3 ,\\
\nn   &&
b_{10,2}^{(7)}= - 2 t u^2 ( (4 m^2 s+5 t^2) D + 2 t^3 u )/s^2D^3 ,\\
\nn   &&
b_{15,2}^{(7)}= 16 ( 2 (t-2 u) D^2 + 2 u^2 (m^2-3 t) D\\
 \nn   && \shift + 3 m^2 t^2 u^2 )/s t u D^2 ,\\
\nn   &&
b_{16,2}^{(7)}= - 2 t (2 D + t u) ( 8 D^2 + 8 s u D - 3 t^2 u^2 )
                                                                 /s^2D^3 ;\\
\nn\\
\nn   &&
b_{1,3}^{(7)}= - 4 ( D^2 - m^2 u D + m^2 T u^2 )/s T D^2,\\
\nn   &&
b_{2,3}^{(7)}= 4 ( 3 T D^2 + m^2 (s t-T u) D
                                       + 2 m^2 t T u^2 )/s t T D^2 ,\\
\nn   &&
b_{3,3}^{(7)}= 4 t u U/sD^2 ,\qquad b_{4,3}^{(7)}= 4 U (D - 2 t u)/sD^2 ,\\
\nn   &&
b_{5,3}^{(7)}= 2 t ( - 8 D^3 - 2 u (t-u) D^2 \\
\nn   && \shift + t u^2 (t-u) D
                                                - m^2 s t u^3 )/s^2D^3 ,\\
\nn   &&
b_{6,3}^{(7)}= 2 t ( - 6 D^3 - 2 u (t-u) D^2 \\
\nn   && \shift + t u^2 (t+2 u) D
                                              - 2 m^2 s t u^3 )/s^2D^3 ,\\
\nn   &&
b_{9,3}^{(7)}= 2 t^2 u ( 4 D^2 - u (t+2 u) D + t u^3 )/s^2D^3 ,\\
\nn   &&
b_{10,3}^{(7)}= 2 t u ( (2 m^2 s (t-u)-3 t^2 u) D
                                              + 2 m^2 s t u^2 )/s^2D^3 ,\\
\nn   &&
b_{15,3}^{(7)}= 16 m^2 ( 2 (t+2 u) D - 3 t u^2 )/s t D^2 ,\\
\nn   &&
b_{16,3}^{(7)}= - 2 t (2 D + t u) (8 D^2 + 2 s u D + 3 t u^3)
                                                                 /s^2D^3 ;\\
\nn\\
\nn   &&
b_{1,4}^{(7)}= 4 t T u/sD^2 ,\\
\nn   &&
b_{2,4}^{(7)}= 4 T ( (3 s-u) D - 2 t^2 u )/s t D^2 ,\\
\nn   &&
b_{3,4}^{(7)}= - 4 ( D^2 - m^2 t D + m^2 t^2 U )/s U D^2 ,\\
\nn   &&
b_{4,4}^{(7)}= 4 ( (m^4-3 t z_u) D - 2 m^4 t u )/s U D^2 ,\\
\nn   &&
b_{5,4}^{(7)}= - 2 t ( (8 m^4 s^2+6 m^2 s t (t-u)+t^2 u^2) D \\
\nn   && \shift
                                                    + t^4 u^2 )/s^2D^3 ,\\
\nn   &&
b_{6,4}^{(7)}= - 2 t ( 6 D^3 + 2 u (t-u) D^2 \\
\nn   && \shift + t^2 u (4 t+3 u) D
                                                  + 2 t^4 u^2 )/s^2D^3 ,\\
\nn   &&
b_{9,4}^{(7)}= 2 t^2 u ( 4 D^2 + t (2 t+3 u) D + t^3 u )/s^2D^3 ,\\
\nn   &&
b_{10,4}^{(7)}= 2 t u ( 2 (t-u) D^2 + t^2 (4 t+3 u) D + 2 t^4 u )
                                                                 /s^2D^3 ,\\
\nn   &&
b_{15,4}^{(7)}= - 16 ( 4 D^2 + 2 t (m^2-2 t) D + 3 m^2 t^3)/s t D^2 ,\\
\nn   &&
b_{16,4}^{(7)}= - 2 t (2 D + t u) ( 8 D^2 - 6 s t D + 3 t^3 u )
                                                                 /s^2D^3 ;\\
\nn\\
\nn   &&
b_{1,5}^{(7)}= - 4 T u^2/sD^2 ,\\
\nn   &&
b_{2,5}^{(7)}= 4 ( D^2 - m^2 u D + 2 t T u^2 )/s t D^2 ,\\
\nn   &&
b_{3,5}^{(7)}= - 4 t^2 U/sD^2 ,\\
\nn   &&
b_{4,5}^{(7)}= 4 ( D^2 - m^2 t D + 2 t^2 u U )/s u D^2 ,\\
\nn   &&
b_{5,5}^{(7)}= - 2 t ( 8 D^3 + 4 t u D^2 - 2 t^2 u^2 D
                                              - m^2 s t^2 u^2 )/s^2D^3 ,\\
\nn   &&
b_{6,5}^{(7)}= 2 t ( - 6 D^3 + 2 (t^2+u^2) D^2 + t^2 u^2 D \\
\nn && \shift                             + 2 t^3 u^3 )/s^2D^3 ,\\
\nn   &&
b_{9,5}^{(7)}= 2 t^2 u ( 4 D^2 + t u D - t^2 u^2 )/s^2D^3 ,\\
\nn   &&
b_{10,5}^{(7)}= - 2 t ( 2 (t^2+u^2) D^2 + t^2 u^2 D + 2 t^3 u^3 )
                                                                 /s^2D^3 ,\\
\nn   &&
b_{15,5}^{(7)}= 16 ( 2 s D^2 - 2 m^2 t u D + 3 m^2 t^2 u^2 )
                                                           /s t u D^2 ,\\
\nn   &&
b_{16,5}^{(7)}= - 2 t (2 D + t u) ( 8 D^2 - 3 t^2 u^2 )/s^2D^3 .
  \ea

\section{}

This Appendix contains the coefficients for the one-loop corrections to the
subprocess $q \bar{q} \rightarrow Q \overline Q$. As regards the box diagram
Fig.~\ref{fig:qqnlo}a we obtain the following coefficients $h$ defined
in Eq.~(\ref{boxq}):
\ba
\nn
& h_1^{(0)} = - 2 T (2/s t - 1/D),   &   \\  \nn
& h_2^{(0)} = 2 (1 + t z_t/\beta^2 D)/s,   &   \\  \nn
& h_4^{(0)} = ( t z_t - s T z_1/D + t z_t/\beta^2 )/D,  &  \\  \nn
& h_6^{(0)} = - 2 t T (1 + s t/D)/D,   &   \\     \nn
& h_7^{(0)} = - t (s^2 T/D - 2 t)/D,   &   \\  \nn
& h_8^{(0)} = ( m^2 s + 2 t^2 + s t^3/D )/D,  &   \\    \nn
& h_{11}^{(0)} = 16 m^2 (T/t - 2 t z_t/s^2\beta^2)/D; &   \\  \nn
\\   \nn
& h_1^{(1)} = - 8 T/s D,      \qquad
  h_2^{(1)} = 8 (t + m^2 z_2/s\beta^2)/s D,  &  \\  \nn
& h_4^{(1)} = - 4 z_t (t^2/D - (1+1/\beta^2)/2)/D,   &  \\  \nn
& h_6^{(1)} = 8 t^2 T/D^2,      \qquad
  h_7^{(1)} = 4 t (2 - t^2/D)/D,  &  \\   \nn
& h_8^{(1)} = 4 s t T/D^2,   \qquad
  h_{11}^{(1)} = - 64 m^2 z_t/s^2\beta^2 D;  &  \\  \nn
\\   \nn
& h_1^{(2)} = z_t/t D,   \qquad
  h_2^{(2)} = - 1/D,          &    \\   \nn
& h_4^{(2)} = s (1 - s t \beta^2/D)/2 D,  &  \\   \nn
& h_6^{(2)} = - t z_1/D^2,   \qquad
  h_7^{(2)} = s t z_2/2 D^2,   &    \\   \nn
& h_8^{(2)} = - s z_1/2 D^2,  \qquad
  h_{11}^{(2)} = - 8 m^2/t D;  &  \\
\\  \nn
& h_1^{(3)} = 0,  \qquad    h_2^{(3)} = 0,  &  \\  \nn
& h_4^{(3)} = s z_t/4 D,   \qquad  h_6^{(3)} = t^2/2 D, &  \\  \nn
& h_7^{(3)} = s t/2 D,   \qquad  h_8^{(3)} = s t/4 D,   \qquad
  h_{11}^{(3)} = 0;  &  \\  \nn
\\   \nn
& h_1^{(4)} = 4 T/t D,   \qquad
  h_2^{(4)} = 4 z_t/s\beta^2 D,  &   \\   \nn
& h_4^{(4)} = 2 (s t z_t/D + 2 m^2 z_2/s\beta^2)/D,   &  \\   \nn
& h_6^{(4)} = h_5^{(4)},  \qquad
  h_7^{(4)} = 2 s t^2/D^2,  &  \\  \nn
& h_8^{(4)} = h_7^{(4)},   \qquad
  h_{11}^{(4)} = - 16 (m^2/t - z_u/s\beta^2)/D;  &   \\  \nn
\\   \nn
& h_1^{(5)} = - 2/D,   \qquad
  h_2^{(5)} = - 2 z_2/s\beta^2 D,    &  \\  \nn
& h_4^{(5)} = s (z_1/D - z_2/s\beta^2)/D,  &  \\  \nn
& h_6^{(5)} = h_5^{(5)},  \qquad
  h_7^{(5)} = s^2 t/D^2,   &  \\  \nn
& h_8^{(5)} = h_7^{(5)}   \qquad
  h_{11}^{(5)} = - 16 z_u/s\beta^2 D;    &  \\  \nn
\\  \nn
& h_i^{(6)} = h_i^{(5)}/2. &
\ea
The values for the other coefficient functions $h_i^{(j)}$ with
$i=3, 5, 9, 10, 12-14$ and arbitrary $j$ are not written out. They can be
inferred from the relations presented in the Eq.~(\ref{hrels}).

The nontrivial coefficients for the second box diagram (\ref{fig:qqnlo}b) are:
\ba
\nn
& h_1^{(0)} = 2 (T/D + 2 (s+U)/s u),  &   \\  \nn
& h_2^{(0)} = - 2 (s/D + 1/s + (2 - u z_u/D)/s \beta^2),   &  \\  \nn
& h_4^{(0)} =  1 - (8 m^2 s + 2 m^2 u - s^2)/D + u t^2
                   (t - u)/D^2    &    \\
                 &  - (2 - u z_u/D)/\beta^2, &  \\   \nn
& h_6^{(0)} = 2 u (m^2 s t/D + m^2 - 2 u)/D,   &  \\    \nn
& h_7^{(0)} = - u (4 s + 2 u - s t^2/D)/D,  &  \\       \nn
& h_8^{(0)} = - 2 + s (m^2-2 u)/D + m^2 s^2 t/D^2,   &  \\    \nn
& h_{11}^{(0)} = - 16 m^2 (s + U + 2 t u z_u/s^2\beta^2)/u D;  &
\ea
The values for the other coefficient functions $h_i^{(0)}$ with
$i=3, 5, 9, 10, 12-15$ are not spelled out. Again they can be
inferred from the relations Eq.~(\ref{hqb1rels}).

Next we write
\ba
\nn
& h_1^{(1)} = 4 (2 m^2 t/u - z_{2u})/s D,  &  \\   \nn
& h_2^{(1)} = 2 z_{2u} (1+1/\beta^2)/s D,  &  \\   \nn
& h_3^{(1)} = - 2 (2 m^2 s^2 \beta^2 + 2 u t z_u + s D)/D^2,  &  \\
\nn
& h_4^{(1)} = 2 (z_{1u} (2 m^2-s)/D + 2 m^2 z_{2u}/s\beta^2)/D,  & \\
\nn
& h_6^{(1)} = 4 u (m^2 s + u z_u)/D^2,  &  \\  \nn
& h_7^{(1)} = 2 (-2 u t^2/D + s+4 u)/D,  &  \\  \nn
& h_8^{(1)} = 2 s (m^2 s + u z_u)/D^2,  &  \\  \nn
& h_{11}^{(1)} = 16 m^2 (3/u - 4 z_u/s^2\beta^2)/D,  & \\  \nn
& h_{14}^{(1)} = 2 u (2 u z_{2u} - s^2 (1+2 \beta^2))/D^2, & \\
\nn
& h_{16}^{(1)} = - 4 z_u/u D,  \qquad   h_{17}^{(1)} = 4/D. &  \\
\\   \nn
& h_1^{(4)} = - 4 U/u D,  \qquad
  h_2^{(4)} = - 4 z_u/s\beta^2 D,  &  \\   \nn
& h_3^{(4)} = - 2 z_{2u} (m^2 s/D - 1/\beta^2)/D,   &  \\  \nn
& h_4^{(4)} = - 2 (s u z_u/D + 2 m^2 z_{2u}/s\beta^2)/D,  & \\  \nn
& h_6^{(4)} = - 4 u^3/D^2,   \qquad
  h_7^{(4)} = 2 s u t/D^2,   &  \\  \nn
& h_8^{(4)} = - 2 s u^2/D^2,   \qquad
  h_{11}^{(4)} = 16 m^2/u D,  &  \\  \nn
& h_{14}^{(4)} = - 2 s u z_{2u}/D^2,   &  \\  \nn
& h_{16}^{(4)} = 4/D,   \qquad   h_{17}^{(4)} = 4 z_{2u}/s\beta^2 D.  &
\ea
The remaining coefficient functions $h_i^{(j)}, j=1,4$ with
$i=5, 9, 10, 12, 13, 15$ can be obtained from
the relations Eq.~(\ref{hqb2rels}).


\end{document}